\documentclass[11pt,a4paper]{article}

\usepackage{graphicx} 
\usepackage{amsmath}  
\usepackage{amsfonts} 

\usepackage[parfill]{parskip}

\usepackage[scale=0.7,vmarginratio={1:2},heightrounded]{geometry}

\usepackage{amsthm} 
\usepackage[affil-it]{authblk} 
\usepackage{amsopn} 
\usepackage{amssymb} 
\usepackage{mathrsfs} 
\usepackage{booktabs} 
\usepackage{natbib} 
\usepackage{tikz} 
\usepackage{algorithm} 
\usepackage{multirow} 

\usepackage{float} 
\usepackage[font={footnotesize}]{caption} 
\usepackage{abstract} 
\usepackage{listings} 
\usepackage{color} 

\usepackage{pdflscape}

\usepackage[pagebackref]{hyperref} 

\newfloat{algorithm}{t}{lop}

\numberwithin{equation}{section}

\newcommand{\RR}{\mathbb{R}}
\newcommand{\PP}{\mathbb{P}}
\newcommand{\EE}{\mathbb{E}}

\newcommand{\de}{\mathrm{d}}
\newcommand{\si}{\Sigma}
\newcommand{\del}{\Delta}

\DeclareSymbolFont{bbold}{U}{bbold}{m}{n}
\DeclareSymbolFontAlphabet{\mathbbold}{bbold}
\newcommand{\ind}{\mathbbold{1}}

\newcommand{\stext}[1]{\text{\scriptsize{#1}}}

\newtheorem{lemma}{Lemma}[section]

\theoremstyle{definition}

\theoremstyle{remark}

\begin{document}
\pagestyle{plain}
\title{
  Residual-Bridge Constructs for Conditioned Diffusions}
  \author{Sean Malory\footnote{Email: s.malory@lancaster.ac.uk} }
\author{Chris Sherlock}
\affil{Department of Mathematics and Statistics \protect\\
          Lancaster University \protect\\
          Lancaster LA1 4YF, UK}
\date{} 

\maketitle

\begin{abstract}
We introduce a new residual-bridge proposal for approximately 
simulating conditioned diffusions. This 
proposal is formed by applying the modified diffusion bridge 
approximation of 
\cite{DGBridge} to the difference between the true diffusion and a 
second, approximate diffusion driven by the same Brownian motion, 
and can be viewed as a natural extension to recent work on 
residual-bridge constructs \citep{Whit}. This new proposal 
attempts to account for volatilities which are not 
constant and can
therefore lead to gains in efficiency over the recently proposed 
residual-bridge constructs in situations where the volatility 
varies considerably, as is often the case for larger 
inter-observation times and for time-inhomogeneous volatilities. These 
potential gains in efficiencies are illustrated via a 
simulation study.
\end{abstract}
   
\section{The Introduction}
\label{sec:Intro}

Diffusions are a flexible class of continuous-time Markov processes 
whose dynamics are completely characterized by specifying an 
instantaneous change in mean (henceforth the drift) and an 
instantaneous variance (henceforth the volatility). This makes them 
a useful class of processes for building rich models and, as such, 
they are utilised in many scientific disciplines, including, but not 
limited to, biology \citep[e.g.][]{Gol11}, finance 
\citep[e.g.][]{Ait07}, and engineering \citep[e.g.][]{Cof04}. In 
biological applications, and, more generally, in applications 
involving reaction networks, diffusions are often used as 
approximate models for the evolution of the numbers of a set of 
species within a reaction network. In particular, the chemical 
Langevin diffusion is often used to approximate the chemical 
master equation \citep[see, for instance,][]{EthKurt, 
Kampen, Wilk, LNA}.

\par A $d$-dimensional diffusion, $X_t$, can be defined as the 
solution to a stochastic differential equation (SDE)
\begin{equation}
\label{eq:SDE}
  \de X_t = \mu(X_t, t, \Theta)\;\de t + 
            \sigma(X_t, t, \Theta)\;\de B_t\;,\quad X_0 = x_0\;,
\end{equation}
where $t\in[0,T]$, $B_t$ is an $r$-dimensional standard Brownian 
motion, and $\Theta\in\RR^{m}$ is a vector of unknown parameters. 
The drift $\mu:\RR^{d}\times[0, T]\times\RR^{m}\rightarrow 
\RR^d$ corresponds to the infinitesimal change in mean, and the 
volatility $\zeta:=\sigma\sigma^T:\RR^{d}\times[0, T]\times
\RR^{m}\rightarrow\RR^{d\times d}$ corresponds to the 
infinitesimal variance in the sense that
\begin{align}
  \label{eq:InfMean}
  \EE(X_{t+\del t}|X_t = x, \Theta = \theta) &= 
                          x + \del t\mu(x, t, \theta) + o(\del t)\;,\\
  \text{Var}(X_{t+\del t}|X_t = x, \Theta = \theta) &=
                          \del t \zeta(x, t, \theta) + o(\del t)\;,
\end{align}
where we write 
$f(\del t) = g(\del t) + o(\del t)$ if and only if
\[\lim\limits_{\del t\downarrow 0}(f(\del t)-g(\del t))/\del t 
  = 0\;.\] 
Both the drift and volatility depend on a vector of unknown 
parameters, $\Theta$, which has a prior density of $p_0(\theta)$. 
These parameters (which drive the evolution of $X_t$) often relate 
to quantities of interest, such as the birth rate of a species, and in 
light of sparse, noisy, and partial observations of the diffusion, 
inference for these parameters, along with paths of the diffusion, can 
theoretically proceed in a Bayesian framework via the particle MCMC 
methodology of \cite{partMCMC}. Such schemes rely on the construction 
of an unbiased approximation to the likelihood of the observations, 
$\pi$, which is typically obtained through an importance-sampling 
and, more generally, particle-filtering approach.

\par Sample paths of the diffusion are infinite-dimensional and 
therefore, in practice, it is necessary to restrict attention to 
the construction of finite-dimensional skeleton paths of the 
diffusion. Moreover, the transition density of a large class of 
diffusions is intractable and exact simulation \citep[e.g.][]{EADiff} 
of a 
skeleton path is impossible for most multivariate diffusions. 
Therefore, for many diffusions, it is necessary to approximate the 
transition density along a fine grid of skeletal points by a 
Gaussian density using an Euler-Maruyama (EM) step.

\par The efficiency of any particle MCMC scheme depends on the 
variability of the importance weights. Hence, the construction of 
proposal densities which are consistent with respect to both the 
observations and the true diffusion is key to designing 
computationally efficient algorithms. The forward simulation (FS) 
proposal of \cite{Ped95} uses the EM approximation to simulate 
skeleton paths between consecutive observations. Such a proposal can 
suffer from poor performance, particularly for informative 
observations, since it simulates paths independently of the 
observations. The modified diffusion bridge (MDB) of 
\cite{DGBridge} overcomes this deficiency by using an EM 
approximation to the transition density between the current point 
of the skeleton and any subsequent point, thus leading to a 
tractable, Gaussian transition density between consecutive points 
of the skeleton given the next observation. However, such a 
proposal performs poorly if sample paths of the diffusion 
exhibit non-linear dynamics as is often the case over relatively 
large inter-observation times. \cite{Lind12} tackles this issue 
by constructing a proposal which is a mixture between the FS 
approach and the MDB approach. The downsides of such a proposal 
are that, firstly, it needs careful tuning, and, secondly, it is 
not clear how the proposal behaves as the mesh of the partition 
tends towards zero. These drawbacks also hold for the proposal of 
\cite{Fearn08} which comprises of a mixture between the FS approach 
and an approach which simulates from the stationary distribution of 
the diffusion (when it exists). \cite{SchauMain} take a different 
approach and consider the form of the SDE satisfied by the diffusion 
conditioned on the next observation; this, in general, has the same 
volatility as the unconditioned diffusion and an extra term in the 
drift \citep[][chapter IV, section 39]{Rogers} which \emph{guides} 
the diffusion towards the observation. This extra term depends on 
the transition density of the unconditioned diffusion and thus, 
typically, needs to be approximated by the transition density of a 
tractable diffusion before forward simulation of a skeleton 
path (via the EM approximation) can proceed. Unfortunately, 
implementing such an approach in a statistically efficient way can 
lead to a computationally expensive algorithm \citep{Whit}.

\par The novel proposal introduced in this paper can be seen as a 
natural extension to the residual-bridge constructs of \cite{Whit} 
who propose improving on the MDB approach by: constructing a 
deterministic path which captures the non-linear dynamics of the 
diffusion, applying the MDB approximation to the residual process 
defined as the difference between the true diffusion and this path, 
and then adding the path back on. An appropriate choice of 
the deterministic path results in a residual whose dynamics are 
more linear and thus a proposal density which is closer to the 
true transition density. It is shown empirically in \cite{Whit} that, 
for several diffusions, this proposal, when implemented within 
a Metropolis-Hastings (MH) importance sampler leads to a larger 
empirical acceptance probability than a MH importance sampler which 
uses either the MDB or the construct introduced by \cite{Lind12} as a 
proposal distribution. Furthermore, this empirical acceptance 
probability is similar to the empirical acceptance probability 
of a MH importance sampler which uses the guided proposals of 
\cite{SchauMain} as a proposal distribution but is 
achieved with a considerably smaller computational cost. However, 
this residual-bridge approach, while accounting for the variability 
in the drift, does not account for the variability in the 
volatility and can, therefore, perform poorly in 
scenarios where the volatility varies substantially. This is 
often the case for larger inter-observation intervals, where the 
diffusion itself moves substantially over the state space, and for 
diffusions whose volatility is time-inhomogeneous. The proposal
introduced in this 
paper generalizes the residual-bridge proposals of 
\cite{Whit} by applying the 
approximation used in the MDB to the difference between the 
true diffusion and a second, carefully chosen, approximate diffusion 
which is coupled with the original diffusion via the same driving 
Brownian motion. By attempting to account for the variability in the
volatility, this new proposal can lead to greater statistical 
efficiency in situations where the volatility varies 
considerably.

\section{Conducting Inference for Diffusions}
\par Let $X_t$ be a $d$-dimensional diffusion satisfying
\eqref{eq:SDE}. Consider the pre-defined sequence of 
times,
\[\{(t_0, \ldots, t_I)\in[0, T]^{I+1}:\;0=:t_0 < 
t_1 < \ldots < t_I := T\}\;.\] 
We have noisy observations,
$(y_{t_1},\ldots, y_{t_I})\in\RR^{r\times I}$, of 
the diffusion at times $(t_1, \ldots, t_I)$ such that, for any 
$i\in\{1, \ldots, I\}$,
\[(Y_{t_i}|X_{t_i} = x) \sim\text{N}(P_ix, \si_{i})\;,\]
where $P_i\in\RR^{r\times d}$, and $\si_i\in\RR^{r\times r}$ is 
symmetric and positive semi-definite. Denote the 
density of the $i$-th observation by $g_i(y_{t_i}|x_{t_i})$ and 
between any two consecutive times, $t_i$ and $t_{i+1}$,
define an equispaced partition, $\mathcal{P}^{i}_{\del t}$, to be 
the set
\[\{(t_{i[0]},\ldots,t_{i[K_i]})\in
[t_i, t_{i+1}]^{K_i+1}:\;t_{i}=:t_{i[0]} < \ldots < t_{i[K_i]} := 
t_{i+1}\}\] 
such that, for all $j\in\{0, \ldots, K_i\}$, 
$t_{i[j]} := t_i + j\del t$ with $\del t>0$ and small. 
For convenience denote any variable $\psi_{t_{i[j]}}$ by $\psi_j^i$ 
with $\psi_{t_{i[0]}}$ denoted by $\psi^i$ so that, for 
instance, $y_{t_{1[0]}} = y^1$ is the first observation, and
$x_{K_I}^{I} = x_T$ is the value of the path at the final time point.
Denote the transition density of the diffusion by
\[f_{\theta}^{s, t}(x|z) := \lim_{\epsilon\downarrow 0}
  \PP(X_t\in [x, x+\epsilon)|X_s = z, \Theta = \theta)/\epsilon^d\;,\]
where
\[[x, x+\epsilon):=\{v\in\RR^d:x_i\leq v_i < x_i + 
  \epsilon\;\text{ for all }\;i\in\{1, \ldots, d\}\}\;.\]
Interest lies in $\pi(\theta, x_{\mathcal{P}_{\del t}}|y^{1:I})$ 
which is the posterior density for $\Theta$ and the 
skeleton path defined at the points of $\mathcal{P}_{\del t}:= 
\mathcal{P}_{\del t}^0 \cup\ldots\cup\mathcal{P}_{\del t}^{I-1}$
and is proportional to
\[
  \underbrace{\pi^{\theta}_0(\theta)}_{\text{Prior for $\theta$}}
  \hspace{-0.5em}\overbrace{\pi^{x^0}_0(x^0)}^{\text{Prior for $x^0$}}
  \hspace{-0.1em}\prod\limits_{i=0}^{I-1}\bigg(\hspace{-0.1em}
  \underbrace{g_{i+1}(y^{i+1}|x^{i+1})}_{
  \text{Observation density}}\hspace{-0.2em}
  \overbrace{\prod\limits_{k=1}^{K_i}f_{\theta}^{t_{i[k-1]}, 
  t_{i[k]}}(x_k^i|x_{k-1}^i)}^{\text{Density of path between
  obs.}}\hspace{-0.3em}\bigg)\;.
\]
The transition density for most diffusions is intractable and
exact simulation techniques \citep[e.g.][]{EADiff} are primarily 
limited 
to diffusions which, under a suitable transformation, have unit 
volatility and, therefore, are typically only applicable to 
one-dimensional diffusions. Hence, for small $\del t>0$, it 
is usual to make the 
following Euler-Maruyama (EM) approximation; 
$f_{\theta}^{t, t+\del t}(x|z)\approx 
\hat{f}_{\theta}^{t, t+\del t}(x|z)$, where we define
\[\hat{f}_{\theta}^{t, t+\del t}(x|z) := 
  \phi(x; z + \del t\mu(z, t, \theta), \del t\zeta(z, t, \theta))\;,
\]
with $\phi(x; m, \Psi)$ denoting the density of a 
Gaussian random variable whose mean 
and variance are $m$ and $\Psi$ respectively. We consider 
the corresponding 
approximate posterior, $\hat{\pi}$, which is proportional to
\[
  \pi^{\theta}_0(\theta)\pi^{x^0}_0(x^0)
  \prod\limits_{i=0}^{I-1} g_{i+1}(y^{i+1}|x^{i+1})
  \prod\limits_{k=1}^{K_i}\hat{f}_{\theta}^{t_{i_{k-1}}, 
  t_{i_k}}(x_k^i|x_{k-1}^i)\;.
\]
This approximation introduces a bias which decreases as $\del t$ 
decreases. Therefore a good proposal must be consistent with the 
diffusion for any small $\del t >0 $. Provided care is taken to 
construct a scheme which does not mix poorly, using, for example, 
ideas in \cite{Gol08}, inference for this 
approximate target can proceed via the particle marginal 
Metropolis-Hastings methodology of \cite{partMCMC}. Such a scheme
involves iterating over different values of $\theta$ and 
through the observations $y^1, \ldots, y^I$. To simplify
notation we henceforth drop $\theta$, and to simplify 
exposition, and the subsequent simulation study, we 
fix $x^0$ and consider only one observation at time $T$.
We emphasise that, from a statistical efficiency point 
of view, \emph{nothing} is lost in making these
simplifications since none of the proposals to be 
discussed in section \ref{sec:diff_prop} depend on more than the 
subsequent observation, hence any difference in 
statistical efficiency for one observation will translate into 
a similar or greater (due to sequential effects) difference in 
statistical efficiency over many observations. However, we also 
emphasise that, by fixing $x^0$, we decrease the computational 
cost of some of the proposals, thereby increasing the apparent
computational efficiency of those proposals. With these
simplifications the approximate target is
\[\hat{\pi}(x_{1:K}|y^1) \propto
  g_1(y^1|x_{K})\prod\limits_{k=1}^{K}
  \hat{f}^{t_{k-1}, t_k}(x_k|x_{k-1})\;,
\]
where, for ease of exposition, we have denoted any variable
$\psi_{j}^0$ by $\psi_{j}$, $K_0$ by $K$, and $t_{0_j}$ by $t_j$. 

\subsection{Proposals Based on Diffusion Bridges}
\label{sec:diff_prop}

\begin{figure}[t]
  \footnotesize
  \centering
  \input{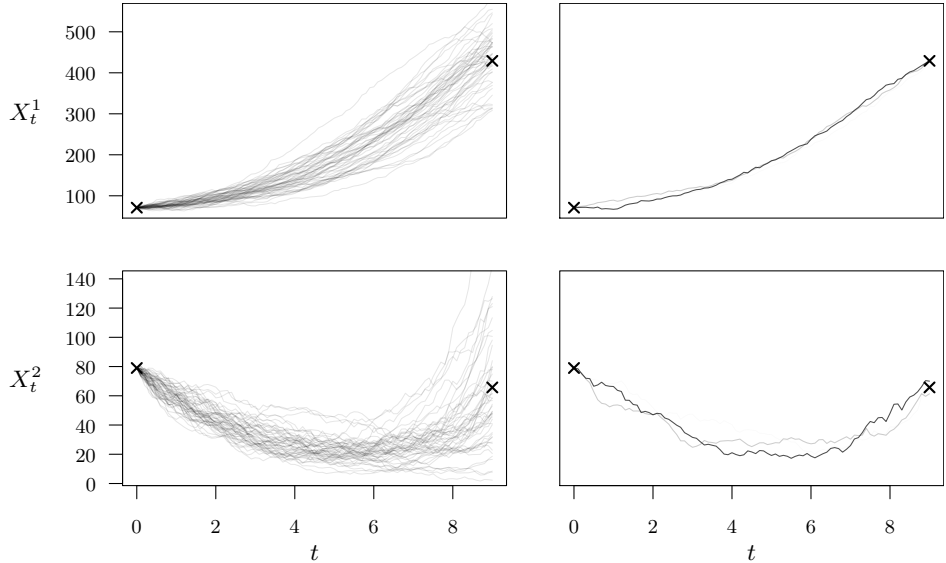}
  \caption{Two sets of plots of fifty paths simulated using the 
           FS approach of \cite{Ped95} on the Lotka-Volterra SDE 
           introduced in subsection \ref{sec:LV}. The plots on the 
           left are the fifty two-dimensional simulated paths with 
           no transparency and the plots on the right are the 
           fifty two-dimensional paths with transparency inversely 
           proportional to their normalised weights so that the 
           path with the largest normalised weight has no transparency 
           and paths with smaller normalised weights are 
           more transparent. The two-dimensional initial condition and 
           observation are illustrated with crosses. It is clear that 
           two of the paths have the highest weight with the 
           other paths having almost zero weight. Moreover, 
           those two paths are precisely the paths whose endpoints 
           lie closest to the observation.}
\label{fig:forward_sim}
\end{figure}

For inexact observations, which are the focus of this paper, the 
particle MCMC methodology requires the sampling of $N$ 
skeleton paths, denoted by $\{x_{1:K}^{(j)}\}_{j=1}^{N}$, 
from a proposal 
$q$ which is close to $\hat{\pi}$ and the calculation of 
the normalised importance weights of the form 
\begin{equation}
  \label{eq:norm_weights}
  \tilde{w}_{j} \propto \hat{\pi}(x_{1:K}^{(j)}|y^1)/
  q(x_{1:K}^{(j)})\;.
\end{equation}
The optimal proposal $q^{\stext{OPT}}\propto \hat{\pi}$ 
results in equal weights, however, for most diffusions such a 
proposal cannot be implemented thus necessitating the need to 
construct proposals which aim to mimic the optimal proposal. The 
forward simulation (FS) approach of \cite{Ped95} uses the proposal
\[q^{\stext{FS}}(x_{1:K}) \propto \prod\limits_{k=1}^{K} 
  \hat{f}^{t_{k-1}, t_k}(x_k|x_{k-1})\;,\]
which leads to weights of the form 
$\tilde{w}_j \propto g(y^1|x_{K}^{(j)})$. Such a proposal produces 
paths which are consistent with the true diffusion but which can 
be inconsistent with the observation since $x_{K}$ is simulated 
irrespective of the value of $y^1$. Therefore, if the noise in the 
observation is small the variability 
of the weights is likely to be large as only a few of the 
simulated endpoints, $x_{K}^{(j)}$ will lie near the 
observation. This phenomena can be seen in figure 
\ref{fig:forward_sim} where we have simulated fifty paths from the 
Lotka-Volterra SDE introduced in subsection \ref{sec:LV} using the FS 
approach of \cite{Ped95}. For illustration purposes we have weighted 
each path under the assumption that the noise in the observation 
is small\footnote{In particular, for all the figures in this section, 
we have assumed that $(Y^1|X_K=x)\sim\text{N}(x, 5I)$.} and have 
plotted the paths twice; the paths on the left 
have no transparency, whereas the paths on the right have been 
plotted with a transparency inversely proportional to their 
normalised weights, $\tilde{w}_j$, so that the path with 
the largest normalised weight has no transparency and the paths 
with smaller normalised weights are more transparent. Thus, 
if there is 
large variability in the weights the number of partially visible 
paths will be small, whereas if there is small variability in the 
weights the number of partially visible paths will be large.
\par
\begin{figure}[t]
  \footnotesize
  \centering
  \input{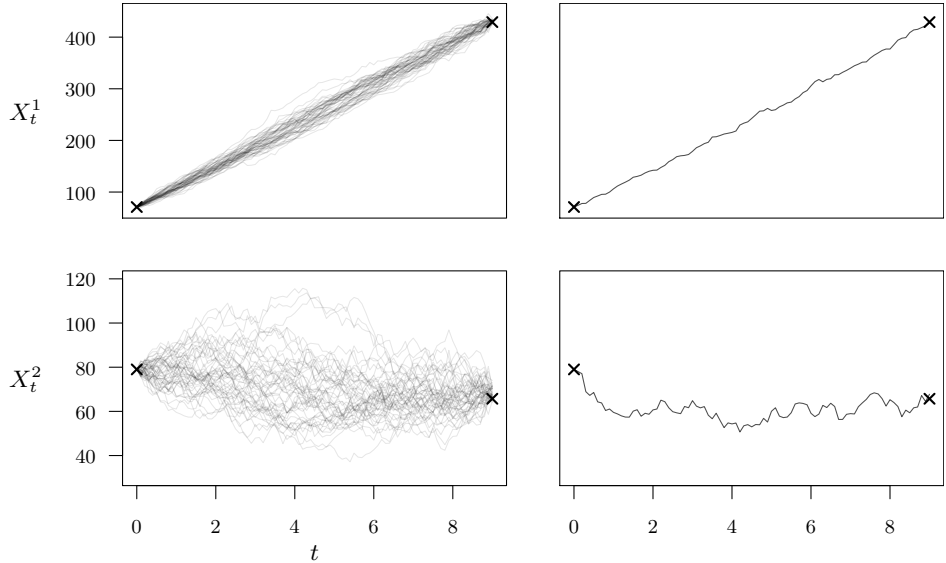}
  \caption{A plot of fifty paths simulated using the MDB of
           \cite{DGBridge} on the Lotka-Volterra SDE introduced in 
           subsection \ref{sec:LV}. As in figure 
           \ref{fig:forward_sim}, the 
           plots on the left are the fifty two-dimensional 
           simulated paths with 
           no transparency and the plots on the right are the 
           fifty two-dimensional paths with transparency inversely 
           proportional to their normalised weights. The 
           two-dimensional initial condition and 
           observation are illustrated with crosses. It can be 
           seen that, even though all of the paths are consistent 
           with the observation, none of the paths are consistent 
           with the dynamics of the true diffusion and hence one 
           of the paths has a much larger weight relative to the 
           other paths.}
\label{fig:mdb}
\end{figure}

 The modified diffusion bridge (MDB) of \cite{DGBridge} overcomes 
the drawback of the FS approach by forming a proposal 
which depends on the observation $y^1$. Specifically, suppose 
that at time $t_k$ we have 
simulated $x_k$. Conditional on this point, form the approximate 
diffusion, $X^{\stext{MDB}}_t$, which satisfies, for $t\in[t_k, T]$,
\begin{equation}
\label{eq:MDBApprox}
  \de X^{\stext{MDB}}_t = \mu(x_k, t_k)\;\de t + 
  \sigma(x_k, t_k)\;\de B_t\;,\quad X^{\stext{MDB}}_{k} = x_k\;.
\end{equation}
This approximation is equivalent to assuming that the EM approximation 
between the current time point and any subsequent time point is 
exact and leads to the following joint distribution for the 
approximate process, $X^{\stext{MDB}}_t$, at the next point of 
the partition and at the observation time;
\[\begin{bmatrix}
	X^{\stext{MDB}}_{k+1} \\
	X^{\stext{MDB}}_{K}
\end{bmatrix}\bigg| (X^{\stext{MDB}}_k = x_k)
\sim\text{N}(m_k^{\stext{MDB}}, \Psi_k^{\stext{MDB}})\;,\]
where
\[
  m_k^{\stext{MDB}}:=
\begin{bmatrix}
	x_k + \del t\mu(x_k, t_k) \\
	x_k + (T-t_k)\mu(x_k, t_k)
\end{bmatrix}\;,\;\Psi_k^{\stext{MDB}} :=
\begin{bmatrix}
	\del t\zeta(x_k, t_k) & \del t\zeta(x_k, t_k) \\
	\del t\zeta(x_k, t_k) & (T-t_k)\zeta(x_k, t_k)
\end{bmatrix}\;.
\]
Consequently, the joint distribution for the approximate process 
at the next point of the partition and the observation, $Y^1$, is 
given by
\[\begin{bmatrix}
	X^{\stext{MDB}}_{k+1} \\
	Y^1
\end{bmatrix}\bigg| (X^{\stext{MDB}}_k = x_k)
\sim\text{N}(\bar{m}_k^{\stext{MDB}}, \bar{\Psi}_k^{\stext{MDB}})\;,\]
where
\[
  \bar{m}_k^{\stext{MDB}}:=
\begin{bmatrix}
  x_k + \del t\mu(x_k, t_k) \\
  P_1x_k + (T-t_k)P_1\mu(x_k, t_k)
\end{bmatrix}\;,\;\bar{\Psi}_k^{\stext{MDB}} :=
\begin{bmatrix}
  \del t\zeta(x_k, t_k) & \del t\zeta(x_k, t_k)P_1^* \\
  \del tP_1\zeta(x_k, t_k) & (T-t_k)P_1\zeta(x_k, t_k)P_1^* + \si_1
\end{bmatrix}\;,
\]
and, in order to avoid confusion with the inter-observation time, 
$T$, we have denoted the transpose of a matrix $A$ by $A^*$.
Standard manipulations for the multivariate normal distribution 
show that
\begin{equation}
\label{eq:MDBprop}
(X^{\stext{MDB}}_{k+1}|X^{\stext{MDB}}_k = x_k, Y^1 = y^1) \sim 
\text{N}(a_k^{\stext{MDB}}, V_k^{\stext{MDB}}) \;,
\end{equation}
where
\begin{alignat*}{2}
  a_k^{\stext{MDB}} & :=\;& &x_k + \del t\mu(x_k, t_k) 
      + \del t\zeta(x_k, t_k)P_1^*((T-t_k)P_1\zeta(x_k,
                                     t_k)P_1^*+\si_1)^{-1}\\
      &     & &\times(y_1 - P_1x_k-(T-t_k)P_1\mu(x_k, t_k))\;,\\
  V_k^{\stext{MDB}} & :=\;& & \del t\zeta(x_k, t_k) - 
                              \del t^2\zeta(x_k, t_k)P_1^* 
         ((T-t_k)P_1\zeta(x_k, t_k)P_1^*+\si_1)^{-1}
                              P_1\zeta(x_k, t_k)\;.
\end{alignat*}
Recall that the approximate process, \eqref{eq:MDBApprox}, is 
equivalent to assuming an EM approximation between the current 
time point and any 
subsequent time point, hence paths simulated using the MDB 
exhibit linear dynamics. Thus, even though paths simulated 
in this way are 
consistent with the observation, they are inconsistent with any 
non-linear dynamics of the true diffusion and, hence, can perform 
poorly in scenarios where the true diffusion exhibits 
non-linear dynamics and particularly, therefore, for relatively larger 
values of $T$. This behaviour, when compared to
figure \ref{fig:forward_sim}, can be seen in figure 
\ref{fig:mdb} where we have simulated fifty paths from the 
Lotka-Volterra SDE introduced in subsection \ref{sec:LV} using the MDB
of \cite{DGBridge}. Again, for illustration purposes, paths have been 
plotted twice; the paths on the left have no transparency, whereas 
the paths on the right have transparency inversely proportional to 
their normalised weights.

\par \cite{Whit} introduce residual-bridge proposals which deal with 
this issue, albeit at a greater computational cost, by constructing 
a deterministic path, $\xi_t$, which captures the non-linear 
dynamics of the true, conditioned diffusion and considering the 
residual, $R_t := X_t - \xi_t$, which satisfies, for $t\in[0, T]$,
\[\de R_t = (\mu(X_t, t) - \xi_t')\de t + \sigma(X_t, t)\de
W_t\;,\quad R_0 = 0\;.\]
If $\xi_t$ accurately captures the non-linear dynamics of the true, 
conditioned diffusion then the residual should exhibit behaviour 
which is more linear, hence applying the MDB to the residual and
adding back $\xi_t$ will result in a proposal which more 
closely resembles the optimal proposal. Suppose, then, 
that at time $t_k$ we have simulated $x_k$. Applying the 
MDB to the residual, $R_t$, gives the 
following joint distribution for the approximate residual process, 
$R^{\stext{RB}}_t$, at the next point of the partition and at the 
observation time;
\[\begin{bmatrix}
	R^{\stext{RB}}_{k+1} \\
	R^{\stext{RB}}_{K}
\end{bmatrix}\bigg| (X^{\stext{RB}}_k = x_k)
\sim\text{N}(\gamma_k^{\stext{RB}}, C_k^{\stext{RB}})\;,\]
where
\begin{align*}
  \gamma_k^{\stext{RB}}:&\!\!=
\begin{bmatrix}
  (x_k-\xi_k) + \del t(\mu(x_k, t_k)-(\xi_{k+1}-\xi_k)/\del t) \\
  (x_k-\xi_k) + (T-t_k)(\mu(x_k, t_k)-(\xi_{k+1}-\xi_k)/\del t)
\end{bmatrix} \\
&\!\!= 
\begin{bmatrix}
  (x_k-\xi_{k+1}) + \del t \mu(x_k, t_k) \\
  (x_k-\xi_k) + (T-t_k)(\mu(x_k, t_k)-(\xi_{k+1}-\xi_k)/\del t)
\end{bmatrix}\;,\\
C_k^{\stext{RB}} :&\!\!=
\begin{bmatrix}
	\del t\zeta(x_k, t_k) & \del t\zeta(x_k, t_k) \\
	\del t\zeta(x_k, t_k) & (T-t_k)\zeta(x_k, t_k)
\end{bmatrix}\;.
\end{align*}
Here $X^{\stext{RB}}_t := R^{\stext{RB}}_t + \xi_t$ denotes the
process which approximates the true process and, as in \cite{Whit}, 
we have approximated $\xi'_k$ via the chord between $(t_k, \xi_k)$ and 
$(t_{k+1}, \xi_{k+1})$. Adding back $\xi_t$ leads to the following 
joint distribution for the approximate process, $X^{\stext{RB}}_t$, 
at the next point of the partition and at the observation time;
\[\begin{bmatrix}
	X^{\stext{RB}}_{k+1} \\
	X^{\stext{RB}}_{K}
\end{bmatrix}\bigg| (X^{\stext{RB}}_k = x_k)
\sim\text{N}\bigg(\begin{bmatrix}
                  m_{k+1}^{\stext{RB}}\\
                  m_{K}^{\stext{RB}}
                  \end{bmatrix}, \Psi_k^{\stext{RB}}\bigg)\;,
\]
where
\begin{alignat*}{2}
  m_{k+1}^{\stext{RB}}&:=\;& & x_k + \del t\mu(x_k, t_k)\;,\\
  m_{K}^{\stext{RB}}&:=\;& & x_k + (\xi_{K}-\xi_k) 
                           +(T-t_k)(\mu(x_k, t_k)-
                           (\xi_{k+1}-\xi_k)/\del t)\;,\\
  \Psi_k^{\stext{RB}} &:=\;& &
\begin{bmatrix}
	\del t\zeta(x_k, t_k) & \del t\zeta(x_k, t_k) \\
	\del t\zeta(x_k, t_k) & (T-t_k)\zeta(x_k, t_k)
\end{bmatrix}\;.
\end{alignat*}
Therefore, the joint distribution for the approximate process 
at the next point of the partition and the observation, $Y^1$, is 
given by
\[\begin{bmatrix}
	X^{\stext{RB}}_{k+1} \\
  Y^1
\end{bmatrix}\bigg| (X^{\stext{RB}}_k = x_k)
\sim\text{N}\bigg(\begin{bmatrix}
                  \bar{m}_{k+1}^{\stext{RB}}\\
                  \bar{m}_{K}^{\stext{RB}}
                \end{bmatrix}, \bar{\Psi}_k^{\stext{RB}}\bigg)\;,
\]
where
\begin{alignat*}{2}
  \bar{m}_{k+1}^{\stext{RB}}&:=\;& & x_k + \del t\mu(x_k, t_k)\;,\\
  \bar{m}_{K}^{\stext{RB}}&:=\;& & P_1x_k + P_1(\xi_{K}-\xi_k) 
                           +(T-t_k)P_1(\mu(x_k, t_k)-
                           (\xi_{k+1}-\xi_k)/\del t)\;,\\
  \bar{\Psi}_k^{\stext{RB}} &:=\;& &
\begin{bmatrix}
	\del t\zeta(x_k, t_k) & \del t\zeta(x_k, t_k)P_1^* \\
	\del tP_1\zeta(x_k, t_k) & (T-t_k)P_1\zeta(x_k, t_k)P_1^*+\si_1
\end{bmatrix}\;.
\end{alignat*}
Standard manipulations for the multivariate normal distribution 
show that
\begin{equation}
\label{eq:RBprop}
(X^{\stext{RB}}_{k+1}|X^{\stext{RB}}_k = x_k, Y^1 = y^1) \sim 
\text{N}(a_k^{\stext{RB}}, V_k^{\stext{RB}}) \;,
\end{equation}
where
\begin{alignat*}{2}
  a_k^{\stext{RB}} & :=\;& &x_k + \del t\mu(x_k, t_k) 
        + \del t\zeta(x_k, t_k)P_1^*((T-t_k)P_1\zeta(x_k,
                                     t_k)P_1^*+\si_1)^{-1}\\
      &     & &\times(y_1 - P_1x_k-P_1(\xi_{K}-\xi_k) 
                                     -(T-t_k)P_1(\mu(x_k,t_k)-
                (\xi_{k+1}-\xi_{k})/\del t))\;,\\
  V_k^{\stext{RB}} & :=\;& & \del t\zeta(x_k, t_k) - 
                              \del t^2\zeta(x_k, t_k)P_1^*
                    ((T-t_k)P_1\zeta(x_k, t_k)P_1^*+\si_1)^{-1}
                              P_1\zeta(x_k, t_k)\;.
\end{alignat*}
The performance of such a proposal clearly hinges on choosing 
a deterministic path $\xi_t$ which has similar dynamics to the true 
diffusion. One natural candidate\footnote{Justified for diffusions 
relating to the chemical Langevin equation by Theorem 2.1 in 
Chapter 11 of \cite{EthKurt}.} for $\xi_t$ is 
constructed by ignoring the volatility in the true diffusion. That 
is, if we let $\xi_t\equiv\eta_t$ be the path obtained by ignoring any 
stochasticity in the evolution of the diffusion, then, from 
\eqref{eq:InfMean} we have that $\eta_t$ satisfies
\[\eta_{t+\del t} = \eta_t + \del t\mu(\eta_t, t) + o(\del t)\;,\]
for any $[t, t+\del t]\subset[0, T)$.
Therefore, $\eta_t$ solves the ordinary differential equation (ODE) 
\begin{equation}
\label{eq:ODE}
\frac{\de\eta_t}{\de t} = \mu(\eta_t, t)\;,\quad\eta_0 = x_0\;,
\end{equation}
over $[0, T]$. We denote the residual-bridge with this choice of 
$\xi_t$ by $\text{RB}^{\stext{ODE}}$. This choice for $\xi_t$ is 
independent of the observation and hence can fail to capture the true 
dynamics of the \emph{conditioned} diffusion, particularly when 
the noise in the observation, $\si_1$, is small and the difference 
between the observation, $y^1$, and the endpoint of the deterministic 
path, $\eta_{K}$, is large. Therefore, paths simulated using this 
proposal can be inconsistent with the \emph{conditioned} diffusion 
when the inter-observation time, $T$, is relatively large, since, 
for larger $T$, the stochasticity in the SDE results in dynamics 
which are inconsistent with $\eta_t$. As suggested by \cite{Whit}, 
this motivates constructing a path $\xi_t$ which is consistent with 
the \emph{conditioned} diffusion by approximating the residual 
$R_t$ with a tractable process 
$\hat{R}_t$ and choosing 
\[\xi_t = \eta_t + \EE(\hat{R}_t|Y^1 = y^1)\;.\]
One choice\footnote{Justified for diffusions relating to the 
chemical Langevin equation by Theorem 2.3 in Chapter 11 of 
\cite{EthKurt}} for the tractable process $\hat{R}_t$ is that given 
by the linear noise approximation (LNA). By Taylor expanding 
around $\eta_t$, defined by \eqref{eq:ODE}, the LNA constructs 
an $\hat{R}_t$ which satisfies a linear SDE, and therefore has
Gaussian transition densities. Indeed, by taking a first-order 
Taylor expansion of the drift and a zeroth-order Taylor expansion 
of the square-root of the volatility, one arrives at an approximate 
process $\hat{R}_t$ which satisfies
\begin{equation}
  \label{eq:LNA}
\de\hat{R}_t = J(\eta_t, t)\hat{R}_t\;\de t + 
               \sigma(\eta_t, t)\;\de B_t\;,\quad\hat{R}_0 = 0\;,
\end{equation}
over the interval $[0, T]$, where $J(\eta_t, t)$ is the $d\times d$ 
Jacobian matrix whose $(i, j)$-th entry is
\[J(\eta_t, t)_{ij}:=
\frac{\partial \mu(x, t)_i}{\partial x_j}\bigg|_{x = \eta_t}\;.\] 
Under this approximation, a tractable form for 
$\EE(\hat{R}_t|Y^1 = y^1)$ is available. The following lemma, 
whose proof is deferred to appendix \ref{app:proof}, derives a 
form which can be implemented in a computationally efficient manner 
because the ODEs that need to be solved do not involve any 
inverses.
\begin{lemma}
  \label{lem:cond_path}
  Let $\hat{R}_t$ be the process which satisfies \eqref{eq:LNA} over
  the interval $[0, T]$ and let $Y_1$ be such that
  \[(Y^1|\hat{R}_T = r) \sim \text{N}(P_1(r+\eta_T), \si_1)\;.\]
  Then
  \[\EE(\hat{R}_t|Y^1 = y^1) = \phi_tG_t^{-1}G_T^{*}P_1^*(P_1\phi_T
                               P_1^*+\si_1)^{-1}(y_1-P_1\eta_T)\;,\]
  where $G_t$ and $\phi_t$ satisfy, for $t\in[0, T]$,  the following 
  ODEs;
  \begin{alignat*}{4}
    \frac{\de G_t}{\de t} & = J(\eta_t, t)G_t\;,\quad & & 
                              G_0 & & = & &\;\;I\;, \\
    \frac{\de\phi_t}{\de t} & = J(\eta_t, t)\phi_t 
                              + \phi_tJ(\eta_t, t)^*
                              + \zeta(\eta_t, t)\;,\quad & & 
                              \phi_0 & &= & &\;\;0\;.
  \end{alignat*}
\end{lemma}
For most diffusions $G_t$ and $\phi_t$ will not be available 
analytically; however, using the Fortran subroutine \texttt{lsoda}
\citep{lsoda}, both can be numerically evaluated in an accurate and 
efficient way at any point of the partition $\mathcal{P}_{\del t}^0$. 
We denote the residual-bridge proposal with this choice of $\xi_t$ 
by $\text{RB}^{\stext{LNA}}$. 
\par Fifty paths simulated from the 
Lotka-Volterra SDE introduced in subsection \ref{sec:LV} using the 
$\text{RB}^{\stext{ODE}}$ and $\text{RB}^{\stext{LNA}}$ 
proposals along with the corresponding deterministic 
paths can be seen in figures 
\ref{fig:residual_bridge_ode} and \ref{fig:residual_bridge_lna} 
respectively. As before, in both figures, the paths have been 
plotted twice; the paths on the left of each figure have no 
transparency, whereas the paths on the right of each figure 
have transparency inversely proportional to their normalised weights.
\begin{figure}[t]
  \footnotesize
  \centering
  \input{residual_bridge_ode_plot.tex}
  \caption{A plot of fifty paths simulated using the 
           $\text{RB}^{\stext{ODE}}$ approach of
           \cite{Whit} from the Lotka-Volterra SDE introduced in 
           subsection \ref{sec:LV}. As with the previous figures the
           plots on the left are the fifty two-dimensional 
           simulated paths with 
           no transparency and the plots on the right are the 
           fifty two-dimensional paths with transparency inversely 
           proportional to their normalised weights.
           The two-dimensional initial condition and
           observation are illustrated with crosses and the 
           deterministic path, $\xi_t=\eta_t$ is plotted with a
           dashed line. When compared 
           with previous figures these paths are more consistent 
           with the true diffusion while still being 
           consistent with the 
           observation, thus the variability in the weights is
           smaller than the previous proposals.}
\label{fig:residual_bridge_ode}
\end{figure}
\begin{figure}[t]
  \footnotesize
  \centering
  \input{residual_bridge_lna_plot.tex}
  \caption{A plot of fifty paths simulated using the  
           $\text{RB}^{\stext{LNA}}$ approach of
           \cite{Whit} from the Lotka-Volterra SDE introduced in 
           subsection \ref{sec:LV}. As with the previous figures the
           plots on the left are the fifty two-dimensional 
           simulated paths with 
           no transparency and the plots on the right are the 
           fifty two-dimensional paths with transparency inversely 
           proportional to their normalised weights. The 
           two-dimensional initial condition and 
           observation are illustrated with crosses and the 
           deterministic path, $\xi_t=\eta_t+\EE(\hat{R}_t|Y^1 = y^1)$ 
           is plotted with a dashed line. It can be seen here that, 
           compared with figure \ref{fig:residual_bridge_ode}, 
           these paths 
           are more consistent with the \emph{conditioned} diffusion 
           and therefore their corresponding weights are less 
           variable.}
\label{fig:residual_bridge_lna}
\end{figure}
Although such approaches account for the non-linear dynamics of the 
diffusion, they still assume a constant volatility over the region 
of interest. This leads to poor performance for diffusions whose 
volatility varies greatly over this interval and in particular, 
therefore, for larger inter-observation times $T$ and for 
diffusions whose volatility is time-inhomogeneous.

\section{New Proposals Based on Diffusion Bridges}
We propose an extension to the approach of \cite{Whit} by 
constructing a \emph{process}, $U_t$, which exhibits 
similar dynamics to the true, conditioned
diffusion and considering the residual process 
$\tilde{R}_t := X_t - U_t$. We begin by constructing a 
deterministic path, $\xi_t$, which exhibits similar dynamics to the 
true diffusion (for instance, the path on which 
$\text{RB}^{\stext{ODE}}$ or $\text{RB}^{\stext{LNA}}$ is based). 
We then use this 
path to construct $U_t$ which is coupled with the true diffusion 
through \emph{the same driving Brownian motion} in such a way that 
paths of $U_t$ exhibit similar stochastic behaviour to paths of $X_t$. 
Specifically, for an arbitrary $u_0$, we define $U_t$ to be the 
process which satisfies 
\[\de U_t = \xi_t'\de t + \sigma(\xi_t, t)\de B_t\;,\quad U_0 = u_0\] 
over the interval $[0, T]$ and which is coupled with $X_t$ 
through \emph{the same driving Brownian motion}, $B_t$. The residual 
process, $\tilde{R}_t$, thus satisfies
\[\de \tilde{R}_t = (\mu(X_t, t) - \xi_t')\de t + (\sigma(X_t, t) -
\sigma(\xi_t, t))\de B_t\;,\]
over the interval $[0, T]$ and with initial condition 
$\tilde{R}_0 = x_0 - u_0$. We proceed by making the same 
approximation used in the MDB: suppose that we have 
simulated $x_k$ at time $t_k$. Form an approximate process, 
$\tilde{R}^{\stext{MDB}}_t$, which satisfies
\[\de \tilde{R}^{\stext{MDB}}_t = (\mu(x_k, t_k) - \xi_{t_k}')\;\de t+ 
(\sigma(x_k, t_k) - \sigma(\xi_k, t_k))\;\de B_t\;,\]
over the interval $[t_k, T]$ and has initial condition 
$\tilde{R}^{\stext{MDB}}_k = x_k - u_k$ (where, as we shall see, 
$u_k$ is the superfluous value of the process $U_t$ at time $t_k$). With this approximation 
we have that, conditional on
having simulated $x_k$ at time $t_k$, the process 
$X^{\overline{\stext{RB}}}_t := U_t + \tilde{R}^{\stext{MDB}}_t$ 
satisfies
\[\de X^{\overline{\stext{RB}}}_t = (\xi_t' + 
\mu(x_k, t_k) - \xi'_{t_k})\;\de t + (\sigma(\xi_t, t) + 
\sigma(x_k, t_k) - \sigma(\xi_k, t_k))\;\de B_t\;,\]
over the interval $[t_k, T]$ and has initial condition 
$X^{\overline{\stext{RB}}}_k = x_k$.
Approximating $\sigma$ by a piecewise constant function on the 
partition $\mathcal{P}_{\del t}$;
\[\sigma(\xi_u, u) = \sum\limits_{k=0}^{K-1}\sigma(\xi_{t_k},
  t_k)\ind_{[t_k, t_{k+1})}(u)\;,\]
where we denote by $\ind_{A}(x)$ the indicator function on the 
set $A$, gives 
the following joint distribution for the approximate process, 
$X^{\overline{\stext{RB}}}$, at the next point of the partition 
and at the observation time;
\[\begin{bmatrix}
    X^{\overline{\stext{RB}}}_{k+1} \\
    X^{\overline{\stext{RB}}}_{K}
  \end{bmatrix}\bigg| (X^{\overline{\stext{RB}}}_k = x_k)
\sim\text{N}\bigg(\begin{bmatrix}
  m_{k+1}^{\overline{\stext{RB}}}\\
  m_{K}^{\overline{\stext{RB}}}
\end{bmatrix}, \Psi_{k}^{\overline{\stext{RB}}}
\bigg)\;,
\]
where
\begin{alignat*}{2}
  m_{k+1}^{\overline{\stext{RB}}}&:=\;& & x_k + \del t\mu(x_k, t_k)\;,\\
  m_{K}^{\overline{\stext{RB}}}&:=\;& & x_k + (\xi_{K}-\xi_k)
                              +(T-t_k)(\mu(x_k, t_k)-
                              (\xi_{k+1}-\xi_k)/\del t)\;,\\
  \Psi_k^{\overline{\stext{RB}}} &:=\; & &
  \begin{bmatrix}
	  \del t\zeta(x_k, t_k) & \del t\zeta(x_k, t_k) \\
    \del t\zeta(x_k, t_k) & \Psi_{K}^{\overline{\stext{RB}}}
  \end{bmatrix}\;, \\
  \Psi_{K}^{\overline{\stext{RB}}} &:=\;& & \del t\zeta(x_k, t_k) 
               + \del t\sum\limits_{j = k+1}^{K-1}
               [(\sigma(\xi_j, t_j) 
               + \sigma(x_k, t_k) - \sigma(\xi_k, t_k))
               (\sigma(\xi_j, t_j) + \sigma(x_k, t_k) - 
             \sigma(\xi_k, t_k))^*]\;.
\end{alignat*}
Thus, using \eqref{eq:RBprop}, we see that
\begin{equation}
\label{eq:Newprop}
(X^{\overline{\stext{RB}}}_{k+1}|X^{\overline{\stext{RB}}}_k = x_k, 
Y_1 = y_1) \sim 
\text{N}(a_k^{\overline{\stext{RB}}}, V_k^{\overline{\stext{RB}}}) \;,
\end{equation}
where
\begin{alignat*}{2}
  a_k^{\overline{\stext{RB}}} & :=\;& &x_k + \del t\mu(x_k, t_k)
            + \del t\zeta(x_k, t_k)P_1^*(P_1
            \Psi_{K}^{\overline{\stext{RB}}}P_1^*+\si_1)^{-1}\\
      &     & &\times(y_1 - P_1x_k-P_1(\xi_{K}-\xi_k)
            -(T-t_k)P_1(\mu(x_k,t_k)-
                (\xi_{k+1}-\xi_{k})/\del t))\;,\\
  V_k^{\overline{\stext{RB}}} & :=\;& & \del t\zeta(x_k, t_k) - 
                              \del t^2\zeta(x_k, t_k)P_1^*
                  (P_1\Psi_{K}^{\overline{\stext{RB}}}P_1^*
                      +\si_1)^{-1}
                      P_1\zeta(x_k, t_k)\;.
\end{alignat*}
This proposal attempts to take into account the variability of 
the drift \emph{and} the variability of the square-root of the 
volatility and therefore should outperform the previous 
residual-bridge construct in scenarios where the square-root of the 
volatility exhibits large variation over the interval $[0, T]$ 
and therefore, in particular, for relatively larger $T$ and 
for volatilities which are time-inhomogeneous. A trade-off 
arises since if the square-root of the volatility varies too much 
then, in many cases of interest, constructing a deterministic 
path, $\xi_t$, which accurately captures the true dynamics of the 
diffusion will be tricky if not impossible. Moreover, in section 
\ref{sec:robust} and in section \ref{sec:discussion} we 
highlight scenarios where our proposed bridge may be 
outperformed by the proposals of \cite{Whit}. However, to 
illustrate why this new residual-bridge 
construct might be preferred over the residual-bridge construct 
of \cite{Whit}, consider constructing bridges to the SDE 
\[\de X_t = \mu(t)\de t + \sigma(t)\de B_t\;,\quad X_0 = x_0\;,\] 
over the interval $[0, T]$. It is clear that if one chooses $\xi_t$ 
to be the solution of the ODE
\[\frac{\de\xi_t}{\de t} = \mu(t)\;,\quad \xi_0 = x_0\;,\]
then, for \emph{any} $\sigma(t)$, this new proposal will, up to a 
discretisation error, simulate exact bridges of $X_t$, whereas, 
the proposal of \cite{Whit} will not. Moreover, the variability in the 
weights corresponding to the residual-bridge proposals of 
\cite{Whit} will increase the more $\sigma(t)$ varies over the 
region of interest.
\par As with the residual-bridge construct of \cite{Whit}, $\xi_t$ can 
be any deterministic path whose dynamics closely match that 
of the true conditioned diffusion. We denote this new proposal, where 
$\xi_t = \eta_t$ with $\eta_t$ defined by \eqref{eq:ODE}, by 
$\overline{\text{RB}}^{\stext{ODE}}$ and, where 
$\xi_t = \eta_t + \EE(\hat{R}_t|Y_1 = y_1)$ with $\hat{R}_t$ 
defined by \eqref{eq:LNA}, by 
$\overline{\text{RB}}^{\stext{LNA}}$. Paths simulated using this 
proposal look very similar to paths simulated using the residual 
bridge proposals of \cite{Whit} as can be seen by comparing
figures \ref{fig:residual_bridge_ode} and 
\ref{fig:residual_bridge_lna}, with 
figures \ref{fig:new_residual_bridge_ode} and 
\ref{fig:new_residual_bridge_lna} which show fifty paths
simulated from the Lotka-Volterra SDE introduced in subsection 
\ref{sec:LV} using the $\overline{\text{RB}}^{\stext{ODE}}$ and 
$\overline{\text{RB}}^{\stext{LNA}}$ 
approaches respectively along with the corresponding deterministic 
paths, $\xi_t$. As throughout this paper, in both figures, the 
paths have been 
plotted twice; the paths on the left of each figure have no 
transparency, whereas the paths on the right of each figure 
have transparency inversely proportional to their normalised weights. 
By comparing the plots on the right of each figure with the 
corresponding plots on the right of figures
\ref{fig:residual_bridge_ode} and \ref{fig:residual_bridge_lna} it 
can be seen that paths simulated using this proposal are more 
consistent with the true \emph{conditioned} diffusion and thus have 
less variable weights.
\begin{figure}[t]
  \footnotesize
  \centering
  \input{new_residual_bridge_ode_plot.tex}
  \caption{A plot of fifty paths simulated using the
           $\overline{\text{RB}}^{\stext{ODE}}$ approach introduced
           in this paper from the Lotka-Volterra SDE introduced in 
           subsection \ref{sec:LV}. As with the previous figures the
           plots on the left are the fifty two-dimensional 
           simulated paths with 
           no transparency and the plots on the right are the 
           fifty two-dimensional paths with transparency inversely 
           proportional to their normalised weights.
           The two-dimensional initial condition and
           observation are illustrated with crosses and the 
           deterministic path, $\xi_t=\eta_t$ is plotted with a 
           dashed line. When compared 
           with figure \ref{fig:residual_bridge_ode} these paths 
           are more consistent with the true \emph{conditioned} 
           diffusion thus the variability in the weights is smaller.}
\label{fig:new_residual_bridge_ode}
\end{figure}
\begin{figure}[t]
  \footnotesize
  \centering
  \input{new_residual_bridge_lna_plot.tex}
  \caption{A plot of fifty paths simulated using the
           $\overline{\text{RB}}^{\stext{LNA}}$ approach introduced
           in this paper from the Lotka-Volterra SDE introduced in 
           subsection \ref{sec:LV}. As with the previous figures the
           plots on the left are the fifty two-dimensional 
           simulated paths with 
           no transparency and the plots on the right are the 
           fifty two-dimensional paths with transparency inversely 
           proportional to their normalised weights.
           The two-dimensional initial condition and
           observation are illustrated with crosses and the 
           deterministic path, 
           $\xi_t=\eta_t+\EE(\hat{R}_t|Y^1 = y^1)$ is plotted with a 
           dashed line. When compared 
           with figure \ref{fig:residual_bridge_lna} these paths 
           are more consistent with the true \emph{conditioned} 
           diffusion thus the variability in the weights is smaller.}
\label{fig:new_residual_bridge_lna}
\end{figure}

\subsection{Computational Considerations}

Comparing the form of $\Psi_{k}^{\overline{\stext{RB}}}$ with the 
form of $\Psi_{k}^{\stext{RB}}$, it can be seen that the 
residual-bridge proposals introduced in this paper have a larger 
computational cost compared to the corresponding 
residual-bridge proposals of \cite{Whit}. We point out, however, 
that this difference in cost 
can be considerably reduced for diffusions relating to the 
chemical Langevin diffusion (see section \ref{sec:Intro} and 
the references therein), where the volatility is of the form
\[\zeta(x, t)=S\Lambda(x, t)^2S^*\;,\] 
where $S\in\RR^{d\times r}$ is a 
constant matrix, and $\Lambda\in\RR^{r\times r}$ is a diagonal matrix.
In this case we can circumvent the calculation of partial sums of 
symmetric matrices of 
size $d\times d$ involved in the calculation of 
$\Psi_{K}^{\overline{\stext{RB}}}$ and instead calculate 
partial sums of vectors of size $r$ by letting
$\sigma(x, t) = S\Lambda(x, t)$ so that  
\[
  \Psi_{K}^{\overline{\stext{RB}}} := \del t\zeta(x_k, t_k) 
              + S\del t\sum\limits_{j = k+1}^{K-1}
               [(\Lambda(\xi_j, t_j) 
               + \Lambda(x_k, t_k) - \Lambda(\xi_k, t_k))
               (\Lambda(\xi_j, t_j) + \Lambda(x_k, t_k) - 
             \Lambda(\xi_k, t_k))^*]S^*\;.
\]
Thus, if $r$ is significantly smaller than $d^2/2$, the computational 
cost of calculating $\Psi_{k}^{\overline{\stext{RB}}}$ can be 
significantly reduced.

\section{A Simulation Study}
\label{sec:sim_study}
In this section we compare the performance of the residual-bridge 
constructs introduced in this paper against the 
corresponding residual-bridge 
constructs of \cite{Whit} and the MDB construct of \cite{DGBridge} 
on two diffusions; the Lotka-Volterra (LV) diffusion 
(\ref{sec:LV}) and a diffusion corresponding to a simple model 
of gene expression (GE, \ref{sec:GE}). 

\subsection{The Lotka-Volterra Diffusion}
\label{sec:LV}

The Lotka-Volterra diffusion \citep[e.g.][]{Wilk} is an approximate model for 
the evolution of the numbers, $X_t = [X_t^{1}, X_t^2]^*$ of two 
species (prey and predators respectively) which are subject to 
three forces; prey reproduce with rate $\theta_1$, predators 
reproduce through eating prey with rate $\theta_2$, and predators 
die with rate $\theta_3$. Such a diffusion satisfies
\[
\begin{bmatrix}
\de X_t^1 \\
\de X_t^2
\end{bmatrix}
=
\begin{bmatrix}
\theta_1X^1_t - \theta_2X^1_tX^2_t\\
\theta_2X^1_tX^2_t - \theta_3X_t^2 
\end{bmatrix}\;\de t +
\begin{bmatrix}
\theta_1X^1_t + \theta_2X^1_tX^2_t & -\theta_2X^1_tX^2_t \\
-\theta_2X^1_tX^2_t & \theta_2X^1_tX^2_t + \theta_3X_t^2
\end{bmatrix}^{1/2}\;\de B_t\;,
\]
where, for a matrix $A$, $A^{1/2}$ denotes any matrix square-root 
so that $(A^{1/2})(A^{1/2})^* = A$.

\subsection{A Diffusion for a Simple Gene Expression Model}
\label{sec:GE}

In this subsection we introduce the diffusion which 
approximates a simple model for 
gene expression \citep[see, for example,][]{Komo09, Sher15}.
This diffusion approximately describes
the evolution of the numbers, $X_t = [R_t, P_t]^*$ of two 
biochemical species (mRNA and protein molecules respectively) 
which are subject to 
three forces; transcription with a time-inhomogeneous rate 
$k_R(t)$, mRNA degradation with rate $\gamma_R$, 
translation with rate $k_P$, and protein degradation with 
rate $\gamma_P$. As in \cite{Komo09, Sher15} we take the rate $k_R(t)$ 
to be of the form 
\[k_R(t) = b_0\exp(-b_1(t-b_2)^2)+b_3\;,\]
so that the complete vector of unknown parameters is 
\[\theta = (\gamma_R, \gamma_P, k_P, b_0, b_1, b_2, b_3)\;.\]
Such a diffusion satisfies
\[
\begin{bmatrix}
\de R_t \\
\de P_t
\end{bmatrix}
=
\begin{bmatrix}
  k_R(t) - \gamma_RR_t\\
  k_PR_t - \gamma_PP_t
\end{bmatrix}\;\de t +
\begin{bmatrix}
  \sqrt{k_R(t) + \gamma_RR_t} & 0 \\
  0 & \sqrt{k_PR_t + \gamma_PP_t}
\end{bmatrix}\;\de B_t\;.
\]

We use the same parameters, 
$\theta$, and initial conditions, $x_0$, as those used in 
\cite{Whit} for the Lotka-Volterra diffusion; 
\[\theta = (\theta_1, \theta_2, \theta_3) = (0.5, 0.0025, 0.3)\;,\quad
x_0 = (71, 79)\;,\]
and we use the following parameters, 
\[\theta = (\gamma_R, \gamma_P, k_P, b_0, b_1, b_2, b_3) = 
  (0.7, 0.72, 3, 80, 0.05, 2, 50)\;,\]
and initial condition $x_0 = (70, 70)$ for the diffusion 
corresponding to the simple model of gene expression. We fix
$\del t$ to be $0.1$ for the LV diffusion and $0.01$ for the GE 
diffusion and choose $10$ equally-spaced values for $T$ 
between; $0$ and $10$ for the LV diffusion and $0$ and $4$ for
the GE diffusion. Moreover, to compare the performance of the 
proposals in challenging scenarios, we choose $P_1 = I$ and 
$\si_1 = 10^{-12}I$ so that the observation; 
\[Y^1|X_{K} = x \sim\text{N}(x, 10^{-12}I)\;,\] 
essentially corresponds to exact observations of the
diffusion\footnote{This small choice of variance in the observation 
  is purely to generate challenging scenarios. In practice, if 
  exact observations of the diffusion were available, the 
  inference procedure would be slightly different 
  \citep[see, for example,][]{Ped95, DGBridge} and is
  considered not here.}
For each value of $T$, we simulated 
$10,000$ values for $Y^1_T$ (where we have emphasised the 
dependence on $T$) using the EM approximation to forward 
simulate values of the path at points of the partition. For each 
collection of $10,000$ values we chose five terminal points for 
$y^1_T$, corresponding to the mean, along with the four $90\%$ 
quantiles along the axes of the principal components. For each 
combination of $(T, y^1_T)$, we ran the MDB of \cite{DGBridge}, 
the residual-bridge construct of \cite{Whit} with the two choices 
for $\xi_t$, $\text{RB}^{\stext{ODE}}$ and 
$\text{RB}^{\stext{LNA}}$, along with the residual-bridge 
construct introduced in this paper with the same two choices for 
$\xi_t$, $\overline{\text{RB}}^{\stext{ODE}}$ and 
$\overline{\text{RB}}^{\stext{LNA}}$. For each of the five 
constructs, we simulated $N=1,000,000$ independent skeleton paths 
and calculated the effective sample size per second (ESS/s) from 
the normalised importance weights \citep{ESS}:
\begin{equation}
  \label{eq:ess_s}  
  \text{ESS/s}\;(\tilde{w}_{1:N}) = 
  \frac{(\tilde{w}_1^2+\ldots + \tilde{w}_{N}^2)^{-1}}{
\text{execution time}}\;.
\end{equation}
To account for variability in the execution time, we calculated 
the average execution time over ten identical runs.
\par For completeness we have included, in appendix
\ref{app:raw_results},
the relative effective sample sizes defined by
\begin{equation}
  \label{eq:rel_ess}
  \text{Rel. ESS}\;(\tilde{w}_{1:N}) = 
  N^{-1}(\tilde{w}_1^2+\ldots + \tilde{w}_{N}^2)^{-1}\;,
\end{equation}
along with the (average) execution times for each proposal and 
for each combination of $(T, y_1^T)$ for the Lotka-Volterra and 
gene-expression diffusions detailed in this section, and
the birth-death diffusion detailed in appendix \ref{app:robust}.

\subsection{Results}
\begin{figure}[t]
  \footnotesize
  \centering
  \input{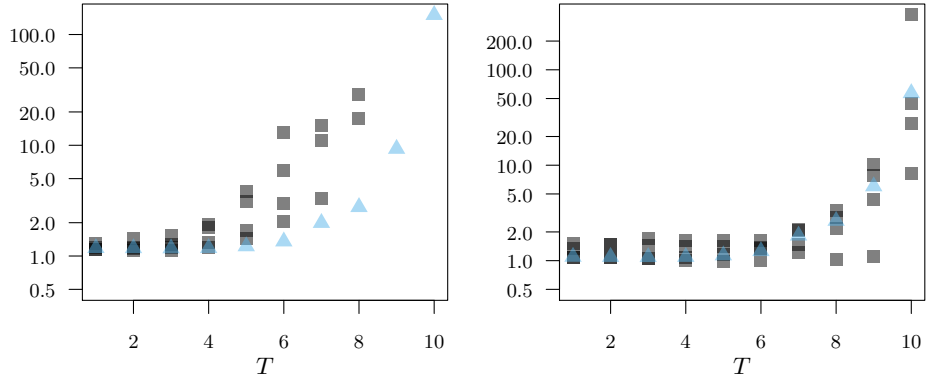}
  \caption{Plots of the comparative effective sample size per 
    second for four pairs of proposals and for a variety of 
    combinations of $(T, y_1^T)$ corresponding to the 
    Lotka-Volterra diffusion. Observations, $y_1^T$, 
    corresponding to the 
    mean of the simulated observations are 
    denoted with blue 
    triangles, whereas the observations corresponding to the 
    four $90\%$ quantiles along the axes of the 
    principal components are denoted with grey boxes.}
\label{fig:results_lv}
\end{figure}

\begin{figure}[t]
  \footnotesize
  \centering
  \input{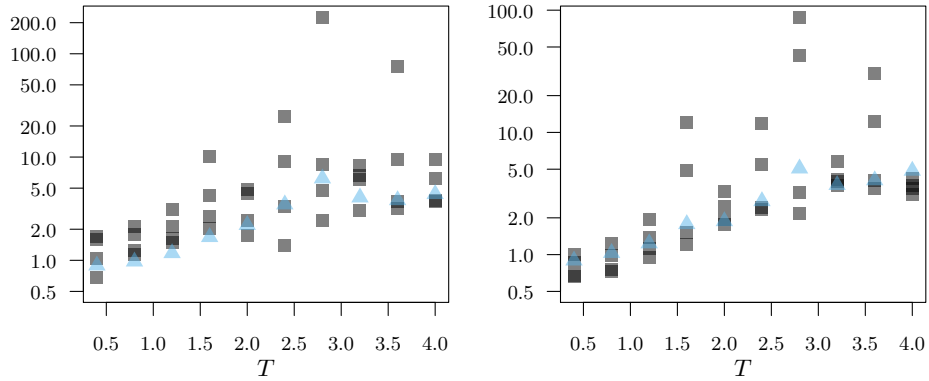}
  \caption{Plots of the comparative effective sample size per 
    second for four pairs of proposals and for a variety of 
    combinations of $(T, y_1^T)$ corresponding to the 
    Lotka-Volterra diffusion. The format is the same as 
  figure \ref{fig:results_ge}.}
\label{fig:results_ge}
\end{figure}

To ease visualisation of comparative performance, figures
\ref{fig:results_lv} and \ref{fig:results_ge}, which illustrate the 
results 
for the LV and GE diffusion respectively, plot, for four pairs of 
proposals, the effective sample size per second for one of the pair of 
proposals relative to the other for each combination of 
$(T, y^1_T)$ for which both proposals had an effective sample 
size of at least one hundred. The four pairs of proposals are chosen 
to approximate the sequential ordering in which the paper has been 
presented. We emphasise that the 
larger the ESS/s the more statistically efficient the proposal is 
for that particular choice of inter-observation time $T$ and 
observation $y^1_T$.
These figures illustrate that the effective sample size 
per second of the
residual-bridge construct introduced in this paper is often similar 
to or larger than the effective sample size per second of the 
corresponding (in the sense of the same deterministic path) 
residual-bridges constructs of \cite{Whit} and, for larger 
inter-observation times, $T$, can exceed it by several 
orders of magnitude.

\subsection{Issues Surrounding Robustness}
\label{sec:robust}
Preserving the discrepancy between the 
square-root volatility at any particular time $t_k$, 
$\sigma(x_k, t_k)$, and the square-root volatility at the same time 
evaluated at the approximating deterministic path, 
$\sigma(\xi_k, t_k)$, can be detrimental to the performance of 
the new residual-bridge proposals when compared to the
proposals of \cite{Whit} in scenarios where preserving such a 
discrepancy leads to a large overestimate/underestimate of the true 
integrated volatility. Therefore it can be argued that 
this new residual-bridge proposal is less robust than the 
residual-bridge proposals of 
\cite{Whit} and care must be taken when implementing such a 
proposal. We illustrate this lack of robustness in appendix
\ref{app:robust} where we compare the performance of the
proposals on a simple one-dimensional diffusion.

\subsection{Issues Surrounding Absolute Continuity}
In this paper we do not prove that the limiting processes 
(as $\del t\downarrow 0$) corresponding to the proposals 
introduced in this paper are absolutely continuous with respect to 
the true conditioned diffusion. Therefore, even though decreasing 
$\del t$ will decrease the bias in our approximate inference scheme, 
this decrease may come at an ever increasing variance, 
as measured by the variability in the weights (i.e. the
effective sample size). However, we provide, through a further 
simulation study detailed in appendix \ref{app:abs_cont}, 
numerical evidence 
suggesting that our proposals are robust to a decreasing $\del t$. 
Specifically, we look at the relative effective sample size 
(\ref{eq:rel_ess}) for the 
new residual-bridge proposals introduced in this paper for 
a variety of inter-observation times $T$, observations $y_1^T$, 
and step-sizes $\del t$ when applied to three diffusions; the 
Lotka-Volterra diffusion and the diffusion for a simple 
model for gene expression introduced in this section, along with 
the birth-death diffusion introduced in appendix \ref{app:robust}. 
We illustrate that the relative effective sample sizes 
are consistent for
\[\del t = 0.1, 0.05, 0.01, 0.005, 0.001\;,\] 
for the birth-death and gene expression diffusions, 
and for 
\[\del t = 0.01, 0.005, 0.001, 0.0005, 0.0001\;,\] 
for the Lotka-Volterra diffusion. This 
consistency of relative effective sample sizes, particularly over 
small values of $\del t$ (which we emphasise are values of 
$\del t$ that border on what is
computationally feasible), strongly suggests that the proposals 
introduced in this paper can be implemented in any 
computationally feasible algorithm (i.e. one that does not 
use a prohibitively small $\del t$) without worrying about the 
effect that decreasing $\del t$ has on the variability of the 
resulting weights.

\section{Discussion of Results}
\label{sec:discussion}
The results of the simulation study illustrated in figures 
\ref{fig:results_lv} and \ref{fig:results_ge} 
show that the performance, in terms of the effective sample 
size per second, of the residual-bridge proposal 
introduced in this paper is often similar to or larger than 
the performance of the residual-bridge proposals of 
\cite{Whit}, and, for larger inter-observation times, $T$, can 
exceed it by several orders of magnitude. Therefore, when 
looked at in conjunction with the analysis of \cite{Whit}, a
particle MCMC scheme which uses this proposal will be more 
efficient than a particle MCMC scheme which uses any
existing proposal and the potential gains in efficiency are 
large. However, as we highlight in appendix \ref{app:robust}, 
there exist some instances where these new
residual-bridge constructs can have a lower (by a factor 
of one half in the worst case found in our simulation study)
effective sample size per second than the corresponding 
constructs of \cite{Whit}. Indeed, one drawback of the proposed 
residual-bridge constructs stems 
from the fact that, at intermediate time points, discrepancies of 
sample paths of the conditional diffusion from the deterministic path, 
$\xi_t$, can be relatively large; preserving the resulting 
discrepancies in the drift and volatility, when for 
$\overline{\text{RB}}^{\stext{LNA}}$ these should be $0$ at time 
$T$, for example, must be sub-optimal. An interpolation scheme 
which is both \emph{justifiable} and \emph{computationally efficient},
however, eludes us.

\par In this paper we have motivated the need for the 
construction of efficient proposals for approximately 
simulating conditioned diffusions over an interval $[0, T]$. 
We have briefly described some of the current proposals used 
in the literature and their drawbacks. We have introduced a 
new residual-bridge proposal and have explained, and demonstrated 
numerically, that such a proposal can often lead to larger 
effective sample sizes for a fixed computational budget, particularly 
for larger 
inter-observation times $T$ and for 
diffusions with volatilities which are time-inhomogeneous. We 
have also highlighted, via a simulation study on a simple 
one-dimensional diffusion, that care needs to be taken when using 
such proposals as they are arguably less robust, across 
different diffusions, than the 
residual-bridge proposals of \cite{Whit}. Further, we have 
provided numerical evidence
which suggests that these new proposals are robust to a decreasing 
step-size $\del t$.
\par All the algorithms in this paper were written in modern Fortran, 
compiled using GNU Fortran (version 4.8.4) from the GNU Compiler 
Collection (http://gcc.gnu.org/) and 
implemented on an Intel Xeon E5-2699 v3 CPU.

\subsection*{Acknowledgements}
S. Malory gratefully acknowledges the support of the EPSRC funded EP/H023151/1 STOR-i centre for doctoral training.

\bibliography{bridges_bibliography.bib}

\appendix

\small
\section{A Proof of Lemma \ref{lem:cond_path}}
\label{app:proof}

\begin{proof}
Define the \emph{generator}, $G_t$, as the solution to 
\[\frac{\de G_t}{\de t} = J(\eta_t, t)G_t\;,\quad G_0 = I\;,\]
over the interval $[0, T]$. Consider the process $G_t^{-1}\hat{R}_t$ 
which satisfies
\begin{align*}
  \de(G_t^{-1}\hat{R}_t) &= \de G_t^{-1}\hat{R}_t 
                           +G_t^{-1}\de\hat{R}_t \\
                         &=-G_t^{-1}\de G_t G_t^{-1}\hat{R}_t 
                           +G_t^{-1}J(\eta_t, t)\hat{R}_t\de t
                           +G_t^{-1}\sigma(\eta_t, t)\de B_t\\
                         &=G_t^{-1}\sigma(\eta_t, t)\de B_t\;.
\end{align*}
Therefore, for any $0 \leq s \leq t \leq T$, $G_t^{-1}\hat{R}_t$ is 
normally distributed with
\begin{equation*}
  \EE(G_t^{-1}\hat{R}_t) = 0\quad,
  \quad\text{Cov}(G_s^{-1}\hat{R}_s,G_t^{-1}\hat{R}_t) =
  \int\limits_{0}^{s}G_u^{-1}\zeta(\eta_u, u)G_u^{-*}\;\de u\;,
\end{equation*}
where $G^{-*}$ is shorthand for $(G^{-1})^*$. Let $\psi_t$ be the 
solution to
\begin{equation}
  \label{eq:InverODE}
  \frac{\de\psi_t}{\de t} = G_t^{-1}\zeta(\eta_t, t)G_t^{-*}\;,
                            \quad\psi_0 = 0\;,
\end{equation}
over the interval $[0, T]$. Then
\[
  \begin{bmatrix}
    \hat{R}_t \\
    Y_1
  \end{bmatrix}
  \sim\text{N}\bigg(
  \begin{bmatrix}
    0\\
    P_1\eta_T
  \end{bmatrix},
  \begin{bmatrix}
    G_t\psi_tG_t^* & G_t\psi_tG_T^*P_1^* \\
    P_1G_T\psi_tG_t^T & P_1G_T\psi_TG_T^*P_1^* + \si_1
  \end{bmatrix}\bigg)\;.
\]
Therefore
\[\EE(\hat{R}_t|Y_1 = y_1) = G_t\psi_tG_T^*P_1^*(
                              P_1G_T\psi_TG_T^*P_1^*+
                              \si_1)^{-1}(y_1-P_1\eta_T)\;.\]
To circumvent the need to calculate $\psi_t$, and therefore 
avoid solving the costly ODE \eqref{eq:InverODE} which contains 
inverses on the right-hand side, we let $\phi_t := G_t\psi_tG_t^{*}$ 
and note that $\phi_t$ solves
\begin{align*}
  \frac{\de \phi_t}{\de t} &= \frac{\de G_t}{\de t}\psi_tG_t^*
                              + G_t\psi_t\frac{\de G_t^*}{\de t}
                              + G_t\frac{\de \psi_t}{\de t}G_t^* \\
                           &= J(\eta_t, t)G_t\psi_tG_t^* + 
                              G_t\psi_tG_t^*J(\eta_t, t)^* +
                              \zeta(\eta_t, t) \\
                           &= J(\eta_t, t)\phi_t 
                              + \phi_tJ(\eta_t, t)^* 
                              +\zeta(\eta_t, t)\;,
\end{align*}
over the interval $[0, T]$ with initial condition $\phi_0 = 0$.
\end{proof}

\section{Issues Surrounding Robustness}
\label{app:robust}
\begin{figure}[t]
  \footnotesize
  \centering
  \input{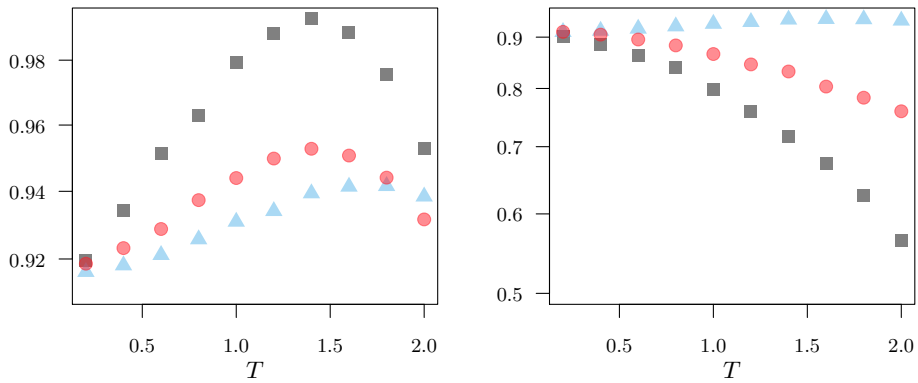}
  \caption{Plots of the comparative effective sample size per 
    second for four pairs of proposals and for a variety of 
    combinations of $(T, y_1^T)$ corresponding to the 
    birth-death diffusion. 
    The observations, $y_1^T$, corresponding to the 
    $50\%$ quantile of the simulated observations are denoted with blue 
    triangles, the observations corresponding to the 
    $5\%$ quantile are denoted with grey boxes, and the observations 
    corresponding to the $95\%$ quantile are denoted with red circles.}
\label{fig:results_bd}
\end{figure}

In this appendix we illustrate, via a simulation study, that 
the new residual-bridge constructs introduced in this paper 
can have a lower effective sample size per second than the 
residual bridge constructs of 
\cite{Whit} and are arguably less robust over different 
diffusions. We will consider a one-dimensional, 
birth-death diffusion $X_t$ \citep{Wilk} which satisfies
\[\de X_t = (\theta_1 - \theta_2)X_t\;\de t+ 
\sqrt{(\theta_1+\theta_2)X_t}\;\de B_t\;,\quad X_0 = x_0\]
over the interval $[0, T]$. This diffusion can be 
considered as an approximate model for the evolution 
of the number, $X_t$, of a species which is subject to two 
forces; births and deaths with rates $\theta_1$ and $\theta_2$ 
respectively. Due to the simplicity of the drift and volatility 
of this diffusion, the term $\eta_t$, defined by 
\eqref{eq:ODE}, along with the terms $G_t$ and $\phi_t$ defined 
in lemma \ref{lem:cond_path} are analytically tractable with
$\eta_t = x_0\exp((\theta_1-\theta_2)t)$, 
$G_t = \exp((\theta_1-\theta_2)t)$, and
\[\phi_t = \frac{(\theta_1 +
\theta_2)}{(\theta_1-\theta_2)}\eta_t
(\exp((\theta_1-\theta_2)t)-1)\;.\]
We conduct a simulation study which mimics the
simulation study of section \ref{sec:sim_study} in order compare the 
performance of the residual-bridge construct introduced in this 
paper against the residual-bridge construct of \cite{Whit} and 
the MDB construct of \cite{DGBridge} on the birth-death 
diffusion. We use the same parameters, 
$\theta$, and initial conditions, $x_0$, as those used in 
\cite{Whit}; $(\theta_1, \theta_2) = (0.1, 0.8)$, 
$x_0 = 50$, so that sample paths of the diffusion 
exhibit exponential decay. We fix $\del t$ 
to be $0.01$ and choose $10$ equally-spaced values for $T$ 
between $0$ and $2$. Moreover, we choose $P_1 = 1$ and 
$\si_1 = 10^{-12}$ so that the observation; 
\[Y^1|X_{K} = x \sim\text{N}(x, 10^{-12})\;,\] 
essentially corresponds to exact observations of the diffusion. 
For each value for $T$, we simulated 
$10,000$ values for $Y^1_T$ (where we have emphasised the 
dependence on $T$) using the EM approximation to forward 
simulate values of the path at points of the partition. For each 
collection of $10,000$ values we chose three terminal points for 
$y^1_T$, corresponding to the $5\%$, $50\%$, and $95\%$ 
quantiles. For each combination of $(T, y^1_T)$, we ran the MDB of 
\cite{DGBridge}, the residual-bridge construct of \cite{Whit} with 
the two choices for $\xi_t$, $\text{RB}^{\stext{ODE}}$ and 
$\text{RB}^{\stext{LNA}}$, along with the residual-bridge 
construct introduced in this paper with the same two choices for 
$\xi_t$, $\overline{\text{RB}}^{\stext{ODE}}$ and 
$\overline{\text{RB}}^{\stext{LNA}}$. For each of the five 
constructs, we simulated $N=1,000,000$ independent skeleton paths 
and calculated, from the normalised importance weights, the effective
sample size per second (ESS/s) as defined by 
\eqref{eq:ess_s}\footnote{As before, to mitigate variability 
  in the execution time, we calculated the average 
  execution time over ten identical runs.}.
As previously, to ease visualisation of comparative performance, 
figure \ref{fig:results_bd} plots, for four 
pairs of proposals, the effective sample size per second for 
one of the pair of proposals relative to the other for 
each combination of $(T, y^1_T)$. The four pairs of 
proposals are chosen to approximate the sequential ordering in 
which the paper has been presented. Again, we emphasise that the 
larger the ESS/s the more statistically efficient the proposal is 
for that particular choice of inter-observation time $T$ and 
observation $y^1_T$.
\par Figure \ref{fig:results_bd} illustrates that the 
effective sample size per second for the new residual-bridge 
construct which uses $\eta_t$, defined by \eqref{eq:ODE}, as 
the deterministic path is similar to, but slightly 
smaller than, due to the increase in computational cost, 
the effective sample size 
per second for the corresponding residual-bridge construct of 
\cite{Whit}. However, the new residual-bridge construct which 
uses $\eta_t+\EE(\hat{R}_t|Y_1 = y_1)$, with $\eta_t$ defined 
by \eqref{eq:ODE} and $\hat{R}_t$ defined by \eqref{eq:LNA}, as the 
deterministic path has an effective sample size per second 
which is significantly worse than the corresponding 
residual-bridge construct of \cite{Whit} for the observations 
corresponding to the $5\%$ and $95\%$ quantiles. This difference is 
particularly large for the observation corresponding to the 
$5\%$ quantile and demonstrates that the performance of the 
new residual-bridge proposals introduced in this paper can be 
worse than that of the residual-bridge proposals of \cite{Whit}, 
thus care needs to be taken when implementing such a proposal. 
We note that, in this example, one can transform the diffusion 
to a diffusion with unit volatility. Specifically, if we let
\[Y_t:=2\sqrt{\frac{X_t}{(\theta_1+\theta_2)}}\;,\]
then $Y_t$ satisfies
\[\de Y_t = \bigg(\frac{(\theta_1-\theta_2)}{2}Y_t -
  \frac{1}{2Y_t}\bigg)\de t + \de B_t\;,\quad Y_0 = 
2\sqrt{\frac{x_0}{(\theta_1+\theta_2)}}\;.\]
As the volatility is constant, applying the residual-bridge 
construct introduced 
in this paper to the transformed diffusion is equivalent 
to applying the residual-bridge construct of 
\cite{Whit} to the transformed diffusion and thus the resulting 
effective sample sizes will be identical (provided, of course, 
the same random numbers are used). However, we emphasise 
that in most cases of practical interest one will not be 
able to transform the diffusion to one of unit volatility and 
therefore care \emph{must} be taken when implementing the 
residual-bridge constructs introduced in this paper.
\par For completeness we have included, in appendix
\ref{app:raw_results},
the relative effective sample sizes (as defined by \eqref{eq:rel_ess})
along with the execution times for each proposal 
and for each combination of
$(T, y_1^T)$ for the birth-death diffusion 
detailed in this appendix, and for the Lotka-Volterra and 
gene-expression diffusions detailed in section \ref{sec:sim_study}.

\section{Issues Surrounding Absolute Continuity}
\label{app:abs_cont}
In this appendix we provide numerical evidence, via 
a simulation study, suggesting 
that the residual-bridge proposals introduced in this 
paper are robust to a decreasing step-size, $\del t$. 
This simulation study will partially extend the studies in 
section \ref{sec:sim_study} and appendix
\ref{app:robust} by considering the two residual-bridge 
constructs introduced in this paper; 
$\overline{\text{RB}}^{\stext{ODE}}$ and 
$\overline{\text{RB}}^{\stext{LNA}}$,
three diffusions; 
birth-death, Lotka-Volterra, and a diffusion 
corresponding to a simple model of gene expression,
and using the same parameters and initial conditions 
as those used in section \ref{sec:sim_study} and 
appendix
\ref{app:robust}. To test the proposals in a 
broad variety of scenarios we chose 
three values for $T$; $(0.2, 1, 2)$ for the 
BD diffusion, $(1, 4, 7)$ for the 
LV diffusion, and $(0.4, 2, 3.6)$ for
the GE diffusion, corresponding 
to a small, medium and large inter-observation interval. 
For each value of $T$ we chose two observations $y_1^T$ 
from the set of observations simulated for the 
simulation studies in section \ref{sec:sim_study} and 
appendix \ref{app:robust}; the centre of the 
simulated observations and one other chosen at random. 
We chose five different values for $\del t$; 
$(0.01, 0.005, 0.001, 0.0005, 0.0001)$ for 
the BD and GE diffusions and 
$(0.1, 0.05, 0.01, 0.005, 0.001)$ for the 
LV diffusion. For each proposal and each combination of 
$(T, y_1^T, \del t)$ we simulated 
$N=1,000,000$ independent skeleton paths 
and calculated the relative effective sample size 
(as defined by \eqref{eq:rel_ess}) from 
the normalised importance weights\footnote{For all of the models 
  and observations the observation variance that was used, 
  $10^{-12}$, is several orders of magnitude smaller than 
  the eigenvalues of the variance matrix at the observation so 
  the empirical evidence of absolute continuity is not affected by 
  this.}.

\begin{landscape}
\begin{table*}[]
\footnotesize
\centering
\caption{A table showing the relative effective 
         sample sizes for $1,000,000$ independent 
         skeleton paths simulated from the two 
         proposals; $\overline{\text{RB}}^{\stext{ODE}}$ and 
         $\overline{\text{RB}}^{\stext{LNA}}$
         for a variety of diffusion 
         models (birth-death, Lotka-Volterra, and 
         gene-expression), inter-observation times (small, 
         medium, and large), step-sizes, and observations (the 
         centre, and one other chosen at random for each 
         combination of $(\text{model}, T, y_1^T)$, but fixed 
         for the different step-sizes). The range of step-sizes 
         are $\del t = 0.01, 0.005, 0.001, 0.0005, 0.0001$ for 
         the BD and GE diffusions, and 
         $\del t = 0.1, 0.05, 0.01, 0.005, 0.001$ for the 
         LV diffusion and the results are displayed in 
         decreasing step-size order. That is, for each 
         group of five results, corresponding to the different 
         values for $\del t$, the effective sample size 
         corresponding to the largest and smallest value for 
         $\del t$ is at the top and bottom of the group 
       respectively.}
\label{tab:abs_cont}
\begin{tabular}{l *{2}c *{2}c *{2}c *{2}c *{2}c *{2}c}
  \toprule
  Proposal & \multicolumn{6}{c}{$\overline{\text{RB}}^{\stext{ODE}}$}
  & \multicolumn{6}{c}{$\overline{\text{RB}}^{\stext{LNA}}$}
  \\
  \midrule
  Diffusion Model & \multicolumn{2}{c}{Birth-Death} &
                  \multicolumn{2}{c}{Lotka-Volterra} &
                \multicolumn{2}{c}{Gene-Expression} & 
                \multicolumn{2}{c}{Birth-Death} &
                \multicolumn{2}{c}{Lotka-Volterra} &
                \multicolumn{2}{c}{Gene-Expression} \\
  \midrule
  Observation & Centre & Other & Centre & Other & Centre & Other
  & Centre & Other & Centre & Other & Centre & Other \\
  \midrule
  \\[0.0015in]
  \multirow{5}{0.6in}{Small $T$}
  & 0.9992 & 0.9990 & 0.9719 & 0.9350 & 0.9370 & 0.8407 & 0.9992 
  & 0.9986 & 0.9716 & 0.9621 & 0.9372 & 0.9054 \\

  & 0.9995 & 0.9991 & 0.9733 & 0.9407 & 0.9408 & 0.8493 & 0.9995 
  & 0.9987 & 0.9731 & 0.9634 & 0.9409 & 0.9083 \\

  & 0.9997 & 0.9992 & 0.9744 & 0.9449 & 0.9441 & 0.8568 & 0.9997 
  & 0.9987 & 0.9744 & 0.9643 & 0.9442 & 0.9107 \\

  & 0.9997 & 0.9992 & 0.9745 & 0.9455 & 0.9444 & 0.8574 & 0.9997 
  & 0.9987 & 0.9745 & 0.9644 & 0.9445 & 0.9108 \\

  & 0.9997 & 0.9992 & 0.9746 & 0.9460 & 0.9446 & 0.8581 & 0.9997 
  & 0.9987 & 0.9746 & 0.9644 & 0.9447 & 0.9112 \\

  \\[0.0015in]
  \midrule
  \\[0.0015in]
  \multirow{5}{0.6in}{Medium $T$}

  & 0.9926 & 0.9878 & 0.6635 & 0.4122 & 0.4289 & 0.2497 & 0.9925 
  & 0.9393 & 0.6574 & 0.6387 & 0.4289 & 0.4029 \\

  & 0.9938 & 0.9890 & 0.6721 & 0.4396 & 0.4355 & 0.2263 & 0.9936 
  & 0.9408 & 0.6694 & 0.6514 & 0.4352 & 0.4118 \\

  & 0.9947 & 0.9898 & 0.6767 & 0.4598 & 0.4469 & 0.2814 & 0.9944 
  & 0.9419 & 0.6765 & 0.6593 & 0.4468 & 0.4190 \\

  & 0.9948 & 0.9899 & 0.6746 & 0.4566 & 0.4442 & 0.2721 & 0.9945 
  & 0.9421 & 0.6749 & 0.6582 & 0.4442 & 0.4107 \\

  & 0.9948 & 0.9900 & 0.6755 & 0.4487 & 0.4372 & 0.2666 & 0.9946 
  & 0.9421 & 0.6757 & 0.6573 & 0.4370 & 0.4077 \\

  \\[0.0015in]
  \midrule
  \\[0.0015in]
  \multirow{5}{0.6in}{Large $T$}

  & 0.9367 & 0.9171 & 0.3379 & 0.0971 & 0.1404 & 0.0740 & 0.9344 
  & 0.7875 & 0.3350 & 0.3172 & 0.1403 & 0.1231 \\

  & 0.9387 & 0.9208 & 0.3678 & 0.0839 & 0.1551 & 0.0806 & 0.9362 
  & 0.7926 & 0.3683 & 0.3289 & 0.1554 & 0.1401 \\

  & 0.9405 & 0.9232 & 0.3688 & 0.0753 & 0.1557 & 0.0905 & 0.9378 
  & 0.7964 & 0.3756 & 0.3377 & 0.1556 & 0.1508 \\

  & 0.9406 & 0.9230 & 0.3709 & 0.0743 & 0.1519 & 0.0785 & 0.9378 
  & 0.7968 & 0.3772 & 0.3381 & 0.1520 & 0.1502 \\

  & 0.9406 & 0.9242 & 0.3664 & 0.0716 & 0.1624 & 0.0776 & 0.9378 
  & 0.7976 & 0.3727 & 0.3375 & 0.1628 & 0.1227 \\

  \\[0.0015in]
  \bottomrule
\end{tabular}
\end{table*}
\end{landscape}

\par Table \ref{tab:abs_cont} shows that the relative 
effective sample size for the proposals introduced in this paper 
are consistent across varying values of $\del t$ for the 
scenarios considered in the simulation study. This therefore 
suggests that such proposals can be implemented without the need 
to consider the effect that decreasing the step-size, $\del t$, has on 
the resulting variability of the weights. Moreover, we stress 
that the smallest $\del t$ considered here is on the border of what is 
computationally feasible, in the sense that any smaller $\del t$, 
with the same inter-observation interval $T$, will lead to an 
algorithm which is prohibitively costly. Therefore,
it can be argued that such proposals are consistent for 
any step-size, $\del t$, that may be used in practice.

\section{Raw Results}
\label{app:raw_results}
\begin{figure}[t]
  \footnotesize
  \centering
  \resizebox{\linewidth}{!}{
          \input{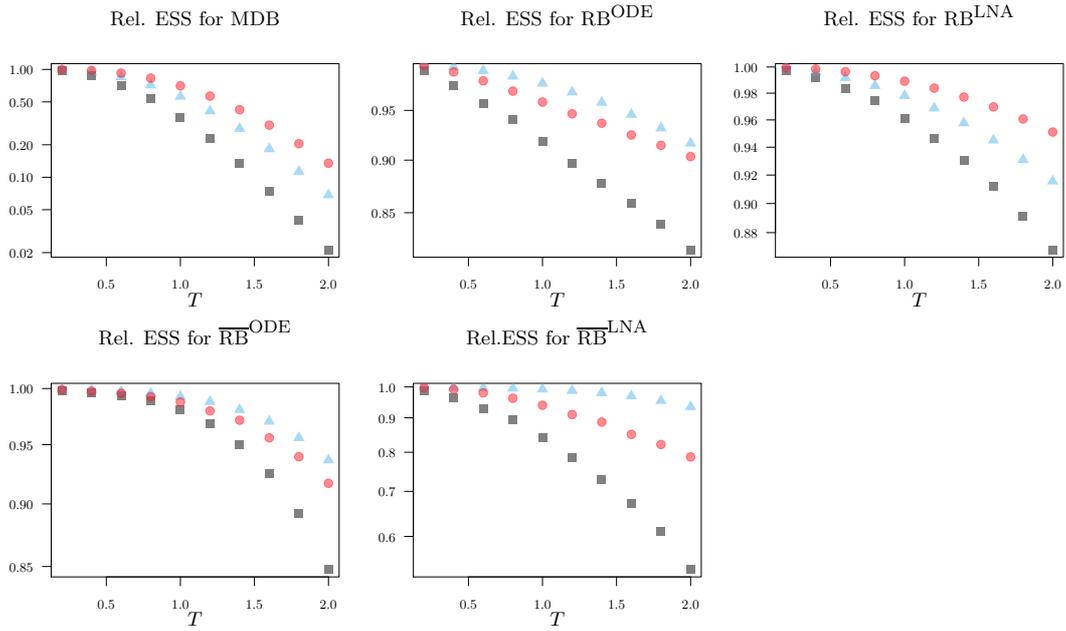}}
  \caption{Plots of the relative effective sample sizes (as defined 
  by \eqref{eq:rel_ess}) for five proposals and for a 
variety of combinations of $(T, y_1^T)$ corresponding to the 
birth-death diffusion. The observations, $y_1^T$, corresponding to 
the 
    $50\%$ quantile of the simulated observations are 
    denoted with blue 
    triangles, the observations corresponding to the 
    $5\%$ quantile are denoted with grey boxes, and the observations 
    corresponding to the $95\%$ quantile are denoted with red circles.}
\label{fig:raw_ess_bd}
\end{figure}

\begin{figure}[t]
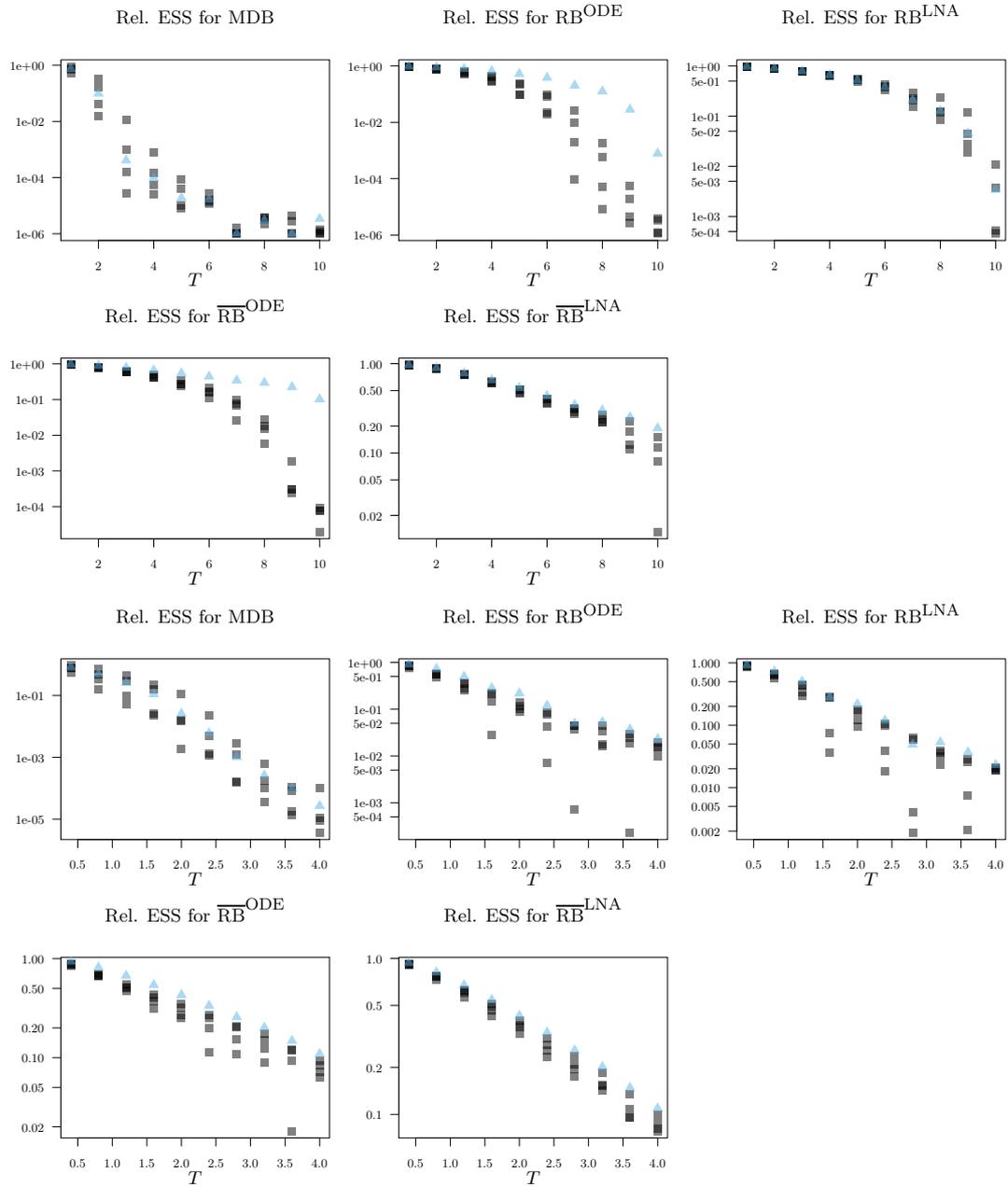

  \footnotesize
  \centering
  \resizebox{\linewidth}{!}{
          \input{lot_vol_raw_ess_plot.tex}}
  \resizebox{\linewidth}{!}{
          \input{hill_raw_ess_plot.tex}}
  \caption{Two sets of five plots of the relative effective sample 
    sizes (as defined 
    by \eqref{eq:rel_ess}) where the top (respectively bottom) five 
    plots are 
    the effective sample sizes for five proposals and for a 
variety of combinations of $(T, y_1^T)$ corresponding to the 
Lotka-Volterra (respectively gene-expression) diffusion. The 
    observations, $y_1^T$, corresponding to the 
    mean of the simulated observations are denoted with blue 
    triangles whereas the observations corresponding to the four 
    $90\%$ quantiles along the axes of the principal components 
    are denoted with grey boxes.}
\label{fig:raw_ess_lv_h}
\end{figure}

\begin{figure}[t]
  \footnotesize
  \centering
  \resizebox{\linewidth}{!}{
        \input{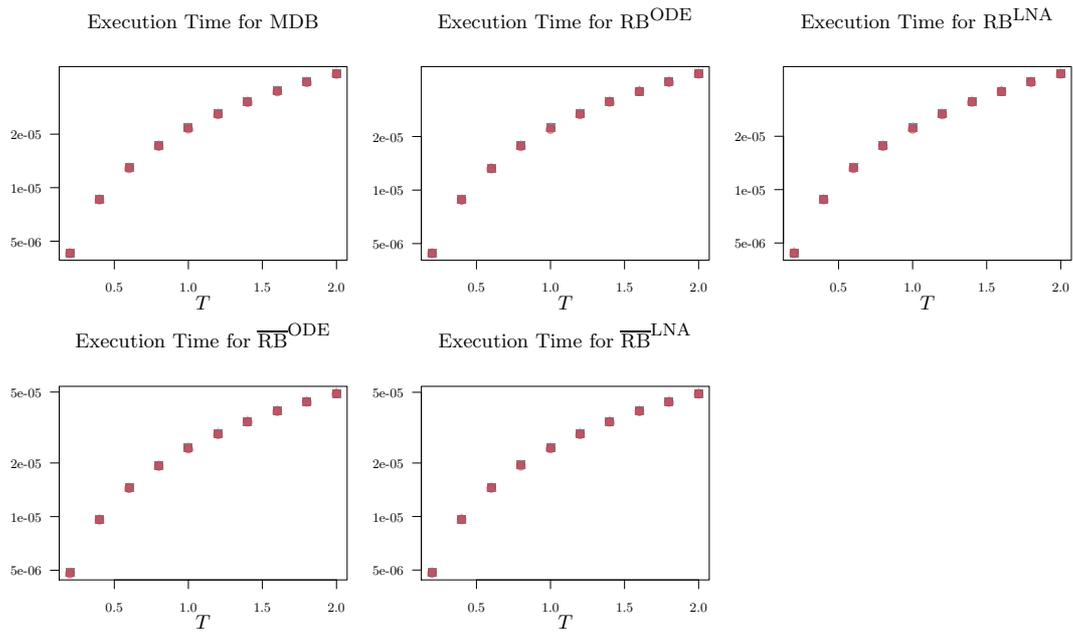}}
  \caption{Plots of the average execution times for five proposals 
    and for a variety of combinations of $(T, y_1^T)$ 
    corresponding to the birth-death diffusion. The observations, 
    $y_1^T$, corresponding to the 
    $50\%$ quantile of the simulated observations are 
    denoted with blue 
    triangles, the observations corresponding to the 
    $5\%$ quantile are denoted with grey boxes, and the observations 
    corresponding to the $95\%$ quantile are denoted with red circles.}
\label{fig:raw_exec_time_bd}
\end{figure}

\begin{figure}[t]
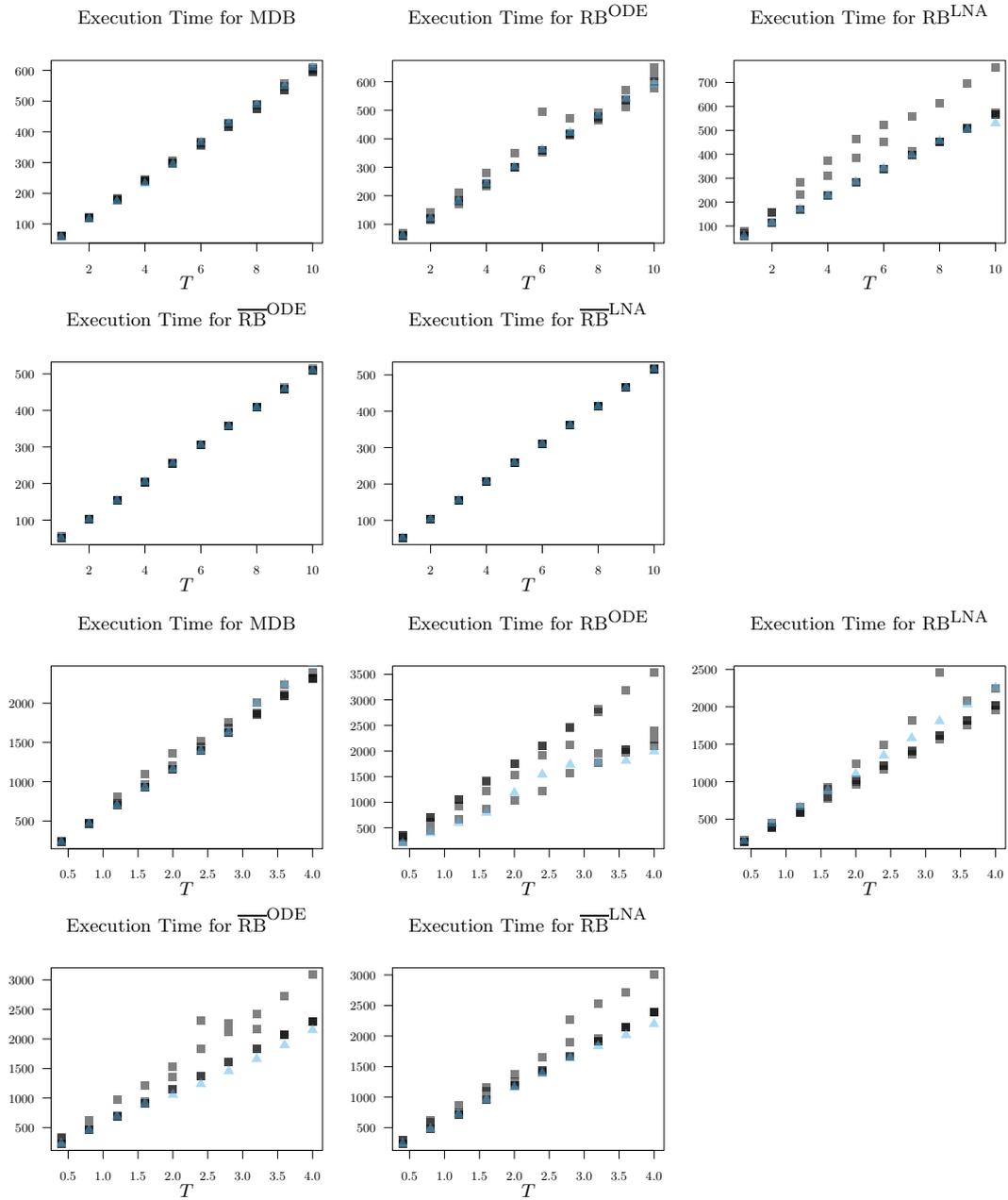

  \footnotesize
  \centering
  \resizebox{\linewidth}{!}{
  \input{lot_vol_raw_exec_time_plot.tex}}
  \resizebox{\linewidth}{!}{
        \input{hill_raw_exec_time_plot.tex}}
  \caption{Two sets of five plots of the average
    execution times where the top (respectively bottom) five 
    plots are 
    the effective sample sizes for five proposals and for a 
variety of combinations of $(T, y_1^T)$ corresponding to the 
Lotka-Volterra (respectively gene-expression) diffusion. The 
    observations, $y_1^T$, corresponding to the 
    mean of the simulated observations are denoted with blue 
    triangles whereas the observations corresponding to the four 
    $90\%$ quantiles along the axes of the principal components 
    are denoted with grey boxes.}
\label{fig:raw_exec_time_lv_h}
\end{figure}

In this appendix we include, for completeness, the raw 
relative effective sample sizes (as defined by \eqref{eq:rel_ess})
and the average execution times for each proposal and each 
combination of
$(T, y_1^T)$ for the Lotka-Volterra and 
gene-expression diffusions detailed in section \ref{sec:sim_study}, 
and for the birth-death diffusion detailed in 
appendix \ref{app:robust}. Recall that, for each 
combination of $(T, y_1^T)$, we 
simulated $1,000,000$ independent skeleton paths using
five different proposals; the MDB of \cite{DGBridge}, 
the residual-bridge proposal of \cite{Whit} with the 
two choices for $\xi_t$, $\text{RB}^{\stext{ODE}}$ and 
$\text{RB}^{\stext{LNA}}$, and the residual-bridge 
proposal introduced in this paper with the same 
two choices for $\xi_t$, $\overline{\text{RB}}^{\stext{ODE}}$ and 
$\overline{\text{RB}}^{\stext{LNA}}$. For each 
proposal and each combination of $(T, y_1^T)$ we calculated 
the normalised weights for each of the $1,000,000$ paths 
according to \eqref{eq:norm_weights} and used these to calculate 
the relative effective sample size (Rel. ESS) defined by
\eqref{eq:rel_ess}. We also noted the average execution time (wall 
time) in seconds over ten identical runs for each algorithm. The 
relative effective sample 
sizes and average execution times can be seen, respectively, in 
figures \ref{fig:raw_ess_bd} 
and \ref{fig:raw_exec_time_bd} for the birth-death
diffusion, and in figures \ref{fig:raw_ess_lv_h} and 
\ref{fig:raw_exec_time_lv_h} for the Lotka-Volterra 
and gene-expression diffusions.

We note that whilst a small relative effective sample size is 
not ideal, indicating as it does a relatively poor proposal, if 
this proposal is the best among its competitors then it is still 
the best option, and, with a large enough absolute effective sample
size, inference which utilises this proposal can still be performed 
accurately.

\end{document}